\documentstyle[times,psfig,color,mathrsfs]{mn}

\newif\ifAMStwofonts
\AMStwofontstrue

\ifoldfss

  \newcommand{\tbd}[1] {\textbf{\color{red}{To be done later}}}
  \ifCUPmtlplainloaded \else
    \NewTextAlphabet{textbfit} {cmbxti10} {}
    \NewTextAlphabet{textbfss} {cmssbx10} {}
    \NewMathAlphabet{mathbfit} {cmbxti10} {} 
    \NewMathAlphabet{mathbfss} {cmssbx10} {} 
  \fi
  \ifAMStwofonts
    \ifCUPmtlplainloaded \else
      \NewSymbolFont{upmath} {eurm10}
      \NewSymbolFont{AMSa} {msam10}
      \NewMathSymbol{\upi}     {0}{upmath}{19}
      \NewMathSymbol{\umu}     {0}{upmath}{16}
      \NewMathSymbol{\upartial}{0}{upmath}{40}
      \NewMathSymbol{\leqslant}{3}{AMSa}{36}
      \NewMathSymbol{\geqslant}{3}{AMSa}{3E}

       \let\le=\leqslant
       
    \fi
  \fi
\fi 

\ifnfssone
  \newmathalphabet{\mathit}
  \addtoversion{normal}{\mathit}{cmr}{m}{it}
  \addtoversion{bold}{\mathit}{cmr}{bx}{it}

  \newcommand{\tbd}[1] {\textbf{\color{red}{To be done later}}}
  \newmathalphabet{\mathbfit} 
  \addtoversion{normal}{\mathbfit}{cmr}{bx}{it}
  \addtoversion{bold}{\mathbfit}{cmr}{bx}{it}
  \newmathalphabet{\mathbfss} 
  \addtoversion{normal}{\mathbfss}{cmss}{bx}{n}
  \addtoversion{bold}{\mathbfss}{cmss}{bx}{n}
  \ifAMStwofonts
    \ifCUPmtlplainloaded \else
      \UseAMStwoboldmath
      \makeatletter
      \new@mathgroup\upmath@group
      \define@mathgroup\mv@normal\upmath@group{eur}{m}{n}
      \define@mathgroup\mv@bold\upmath@group{eur}{b}{n}
      \edef\UPM{\hexnumber\upmath@group}
      \new@mathgroup\amsa@group
      \define@mathgroup\mv@normal\amsa@group{msa}{m}{n}
      \define@mathgroup\mv@bold\amsa@group{msa}{m}{n}
      \edef\AMSa{\hexnumber\amsa@group}
      \makeatother
      \mathchardef\upi="0\UPM19
      \mathchardef\umu="0\UPM16
      \mathchardef\upartial="0\UPM40
      \mathchardef\leqslant="3\AMSa36
      \mathchardef\geqslant="3\AMSa3E

       \let\le=\leqslant

    \fi
  \fi
\fi 

\ifnfsstwo

  \newcommand{\tbd}[1] {\textbf{\color{red}{To be done later}}}
  \DeclareMathAlphabet{\mathbfit}{OT1}{cmr}{bx}{it}
  \SetMathAlphabet\mathbfit{bold}{OT1}{cmr}{bx}{it}
  \DeclareMathAlphabet{\mathbfss}{OT1}{cmss}{bx}{n}
  \SetMathAlphabet\mathbfss{bold}{OT1}{cmss}{bx}{n}
  \ifAMStwofonts
    \ifCUPmtlplainloaded \else
      \DeclareSymbolFont{UPM}{U}{eur}{m}{n}
      \SetSymbolFont{UPM}{bold}{U}{eur}{b}{n}
      \DeclareSymbolFont{AMSa}{U}{msa}{m}{n}
      \DeclareMathSymbol{\upi}{0}{UPM}{"19}
      \DeclareMathSymbol{\umu}{0}{UPM}{"16}
      \DeclareMathSymbol{\upartial}{0}{UPM}{"40}
      \DeclareMathSymbol{\leqslant}{3}{AMSa}{"36}
      \DeclareMathSymbol{\geqslant}{3}{AMSa}{"3E}

       \let\le=\leqslant

    \fi
  \fi
\fi 

\ifCUPmtlplainloaded \else
  \ifAMStwofonts \else 
    \def\upi{\pi}
    \def\umu{\mu}
    \def\upartial{\partial}
  \fi
\fi

   \title[Local Group dwarfs in a cosmological context -- I]{Chemical evolution 
     of Local Group dwarf galaxies in a cosmological context -- I. A new 
     modelling approach and its application to the Sculptor dwarf spheroidal 
     galaxy}
   \author[D.~Romano and E.~Starkenburg]{Donatella Romano$^{1}$\thanks{E-mail: 
       donatella.romano@oabo.inaf.it (DR); else@uvic.ca (ES)} and Else 
     Starkenburg$^{2}$\footnotemark[1]\thanks{CIfAR Junior Fellow and CITA 
       National Fellow.}\\
     $^{1}$INAF, Astronomical Observatory of Bologna, Via Ranzani 1, 40127 
     Bologna, Italy\\
     $^{2}$Department of Physics and Astronomy, University of Victoria, PO Box 
     3055 STN CSC, Victoria, BC, V8W 3P6, Canada}

\begin{document}

     \date{Accepted 2013 June 7. Received 2013 June 7; in original form 2013 January 8}

     \pagerange{\pageref{firstpage}--\pageref{lastpage}} \pubyear{2013}

     \maketitle

     \label{firstpage}


   \begin{abstract}
     We present a new approach for chemical evolution modelling, specifically 
     designed to investigate the chemical properties of dwarf galaxies in a 
     full cosmological framework. In particular, we focus on the Sculptor dwarf 
     spheroidal galaxy, for which a wealth of observational data exists, as a 
     test bed for our model. We select four candidate Sculptor-like galaxies 
     from the satellite galaxy catalogue generated by implementation of a 
     version of the Munich semi-analytic model for galaxy formation on the 
     level~2 Aquarius dark matter simulations and use the mass assembly and 
     star formation histories predicted for these four systems as an input for 
     the chemical evolution code. We follow explicitly the evolution of several 
     chemical elements, both in the cold gas out of which the stars form and in 
     the hot medium residing in the halo. We take into account in detail the 
     lifetimes of stars of different initial masses, the distribution of the 
     delay times for type Ia supernova explosions and the dependency of the 
     stellar yields from the initial metallicity of the stars. We allow large 
     fractions of metals to be deposited into the hot phase, either directly as 
     stars die or through reheated gas flows powered by supernova explosions. 
     We find that, in order to reproduce both the observed metallicity 
     distribution function and the observed abundance ratios of long-lived 
     stars of Sculptor, large fractions of the reheated metals must never 
     re-enter regions of active star formation. With this prescription, all the 
     four analogues to the Sculptor dwarf spheroidal galaxy extracted from the 
     simulated satellites catalogue on the basis of luminosity and stellar 
     population ages are found to reasonably match the detailed chemical 
     properties of real Sculptor stars. However, all model galaxies do severely 
     underestimate the fraction of very metal-poor stars observed in Sculptor. 
     Our analysis thus sets further constraints on the semi-analytical models 
     and, at large, on possible metal enrichment scenarios for the Sculptor 
     dwarf spheroidal galaxy.
   \end{abstract}



   \section{Background and motivation}
   \label{sec:int}

   Understanding how galaxies form and evolve is a major challenge to modern 
   astrophysics. Crucial information on the sequence of events by which a 
   stellar system took shape is encoded in the chemical composition of its 
   stars. To fully exploit this information, large spectroscopic surveys of 
   stars in the Milky Way and its neighbouring systems have been conceived, 
   aimed at providing a complete picture of the chemical enrichment of the 
   interstellar medium (ISM) back to the earliest epochs. At the same time, 
   theoreticians have started to develop more sophisticated galaxy formation 
   models (see, e.g., Font et al. 2011; Brook et al. 2012; Grieco et al. 2012; 
   Pilkington et al. 2012; Revaz \& Jablonka 2012, among others) to account for 
   the unexpected features that are emerging from the data. In this context, 
   studies of Local Group dwarf galaxies play a key role. The reason for this 
   is, at least, threefold. Firstly, dwarf galaxies are the most common type of 
   galaxy in the (local) Universe. As such, they bear the potential to study 
   different star formation histories (SFHs) and their relation with the 
   environment in a large, statistically significant sample of objects of the 
   same class. Secondly, our capability to resolve Local Group dwarfs' stellar 
   populations makes them ideal test beds for theories of galaxy formation and 
   evolution on small scales (see next paragraphs). Thirdly, small systems are 
   the fundamental building blocks for the assembly of larger galaxies in Cold 
   Dark Matter (CDM) cosmological models (White \& Rees 1978; White \& Frenk 
   1991). It has been well established on the basis of chemical arguments (e.g. 
   Shetrone et al. 2001) that the Galactic halo could not form through merging 
   of low-mass galaxies resembling the dwarf spheroidals orbiting the Milky Way 
   \emph{today}. However, very and extremely metal-poor stars belonging to 
   different environments display more similar chemical abundance patterns (of, 
   in particular, $\alpha$ and heavy elements; e.g. Tolstoy et al. 2009). This 
   has renewed the interest in the connection between \emph{ancient} dwarf 
   galaxies and the primeval building blocks of the Galactic halo.

   Both classical bright and faint Local Group dwarf galaxies are close enough 
   to allow their chemical enrichment and star formation histories to be 
   derived with great precision from high-quality measurements of individual 
   stars (Tolstoy et al. 2009, and references therein), which provides tight 
   constraints on simulations of the chemical and dynamical evolution of such 
   systems. In particular, the adoption of the `true' SFH and stellar initial 
   mass function (IMF) derived from analyses of the colour-magnitude diagrams 
   (CMDs) allows us to remove some important parameters of chemical evolution 
   and leads to a more sound modelling of specific objects (Carigi et al. 2002; 
   Lanfranchi \& Matteucci 2003, 2004; Romano et al. 2006).

   Dwarf galaxies should be prone to considerable loss of matter through 
   galactic winds originating from multiple SN explosions, because of their 
   shallow potential wells. Galactic winds have been discussed in theory 
   (Larson 1974; Saito 1979; Matteucci \& Chiosi 1983; Matteucci \& Tosi 1985; 
   Dekel \& Silk 1986; Vader 1987; Pilyugin 1993, among others) long before the 
   observational evidence for them in dwarfs became clear-cut (Meurer et al. 
   1992; Martin 1999; Heckman et al. 2001; Martin et al. 2002; see also 
   Veilleux et al. 2005, and references therein). Nowadays, galactic-scale 
   outflows are often invoked as the most suited explanation for the low mass 
   densities, metallicities and detailed chemical abundance ratios of 
   star-forming dwarfs (e.g. Yin et al. 2011). Yet, a sound theory of galactic 
   winds is missing. In particular, it is still debated whether the SN ejecta 
   can leave the galaxy definitively or cool down and be reaccreted; moreover, 
   it is still unclear which fraction of the ambient gas is entrained in the 
   outflow and how metals are loaded (Silich \& Tenorio-Tagle 1998; D'Ercole \& 
   Brighenti 1999; Mac Low \& Ferrara 1999; Ferrara \& Tolstoy 2000; 
   Tenorio-Tagle et al. 2007; Recchi \& Hensler 2013; see also Heckman et al. 
   2001; Summers et al. 2001, 2003; Martin et al. 2002; Cannon et al. 2004; 
   Oppenheimer \& Dav{\' e} 2006; Peeples \& Shankar 2011).

   When modelling the chemical evolution of dwarf galaxies with estimated SFHs, 
   the final fate and chemical composition of the outflow are only one of the 
   major sources of uncertainty. We still do not know when or how the gas is 
   accreted, or how efficiently it is turned into stars. Recent or ongoing star 
   formation activity in several dwarf irregular galaxies (dIrrs) and blue 
   compact dwarfs (BCDs) is clearly associated to accretion of neutral gas, 
   either through infall or gas-rich body encounters (e.g. Putman et al. 1998; 
   Stil \& Israel 2002; Pustilnik et al. 2003; Kobulnicky \& Skillman 2008). 
   Starbursting dwarfs with no obvious external trigger mechanism could still 
   have faint undetected companions (see Hunter \& Elmegreen 2004, and 
   references therein). On the other hand, the presence of neutral gas does not 
   assure ongoing star formation by itself. Photoionization, heating from the 
   cosmic ultraviolet (UV) background, tidal interactions and ram pressure 
   stripping are further processes that may leave an imprint on the observed 
   properties of dwarf galaxies and should be accounted for in the models.

   In dealing with the above issues, classical chemical evolution models for 
   dwarf galaxies tend to keep things simple. They do not take into account 
   reionization, tidal interactions or stripping of gas and/or stars, and 
   introduce a number of free parameters and simplifying assumptions about the 
   dark matter mass and distribution, the stellar feedback efficiency and the 
   history of mass assembly. In particular, a time-decaying gas infall rate is 
   usually assumed (e.g. Chiosi \& Matteucci 1982; Matteucci \& Chiosi 1983; 
   Bradamante et al. 1998; Mouhcine \& Contini 2002; Yin et al. 2011), 
   following dynamical studies of the collapse of protogalaxies dating to 
   Larson (1976). An infall rate increasing with time, simulating the late 
   accretion of gaseous lumps, was adopted by Romano et al. (2006) for the 
   specific case of NGC\,1569, an exceptionally active nearby dIrr, and shown 
   to nicely reproduce the observed properties of that galaxy. However, no 
   physical explanation was given for the assumed infall law. Models without 
   inflow have also been proposed in the literature (e.g. Carigi et al. 1995).

   Cosmological simulations provide crucial information on the mass assembly 
   history of galaxies and, hence, a tempting framework for chemical evolution 
   studies, since they remove the need for some \emph{ad hoc} prescriptions. In 
   turn, the chemical evolution models have the capability to screen many 
   different realizations for a given object ---or class of objects--- and look 
   for the evolutionary path that maximizes the agreement between the observed 
   and predicted detailed chemical properties. Chemical evolution studies thus 
   offer a way to further constrain the parameters entering \emph{ab initio} 
   galaxy formation models. So far, a number of studies has been devoted to the 
   chemical evolution of galaxies within a hierarchical clustering scheme 
   (Thomas 1999; Nagashima et al. 2005; Nagashima \& Okamoto 2006; Pipino et 
   al. 2009; Arrigoni et al. 2010; Rahimi et al. 2011; Brook et al. 2012), with 
   some attempts to deal with local dwarfs (Salvadori et al. 2008; Calura \& 
   Menci 2009; Okamoto et al. 2010; Sawala et al. 2010; Pilkington et al. 2012; 
   Revaz \& Jablonka 2012). Most of these works, however, consider only a few 
   chemical species; moreover, the contribution of the low- and 
   intermediate-mass stars (LIMS) is often neglected: this prevents a proper 
   treatment of elements such as He, C and N, that are precious diagnostics of 
   chemical evolution, especially in dIrrs and BCDs (e.g. James et al. 2010), 
   as well as important gas coolants (Wiersma et al. 2009). In some cases, type 
   Ia SNe (SNeIa) ---the major iron producers--- are not included in the models.

   In this work, we present a new chemical evolution model, specifically 
   designed to follow the evolution of the abundances of several elements (H, 
   D, He, Li, C, N, O, Na, Mg, Al, Si, S, Ca, Sc, Ti, Cr, Mn, Co, Ni, Fe, Cu, 
   Zn) in the hot and cold gas phases of dwarf galaxies in a cosmological 
   context. The lifetimes of stars of different initial masses and the 
   distribution of the delay times for SNIa explosions are taken into account 
   in detail, as is the dependency of the yields on the initial metallicity of 
   the stars. We choose the Sculptor dwarf spheroidal galaxy (dSph) as a test 
   bed for our model, because of the conspicuous body of high-quality 
   observational data available in this case to calibrate the model parameters. 
   First, four Milky Way satellites resembling Sculptor in luminosity and SFH 
   are identified in the satellite catalogue generated by Starkenburg et al. 
   (2013a) through the implementation of a version of the Munich semi-analytic 
   model (SAM) for structure formation (De Lucia \& Blaizot 2007; De Lucia et 
   al. 2008; Li et al. 2009, 2010) on the high resolution level~2 Aquarius dark 
   matter simulations (Springel et al. 2008a,b). Then, the detailed chemical 
   properties of the four Sculptor candidates are computed from the full 
   cosmological mass assembly histories and including the response to a variety 
   of internal and external physical processes, such as photoionization, 
   heating from the cosmic UV background, star formation, SN feedback, tidal 
   interactions and ram pressure stripping.

   This paper is organized as follows. Section~2 summarizes seven decades of 
   investigation of the Sculptor dSph, highlighting the latest results. 
   Section~3 contains a brief description of the adopted cosmological 
   simulations and semi-analytical model, along with a detailed description of 
   the post-processing chemical evolution code that we have developed. 
   Section~4 follows with a presentation of the results concerning the detailed 
   chemical properties of our four Sculptor candidates. We furthermore point 
   out the dependence of the results on model parameters in Section~4. The 
   strengths and shortcomings of our approach are discussed in Section~5, also 
   in comparison with previous work. We conclude with a summary and prospects 
   for future work in Section~6.

   \section{The Sculptor dwarf spheroidal galaxy}
   \label{sec:data}

   Since its discovery late in the thirties (Shapley 1938), the Sculptor dwarf 
   spheroidal galaxy (dSph) has been the subject of extensive investigation. 
   Sculptor is a relatively faint ($M_V \approx -$11.1; Irwin \& Hatzidimitriou 
   1995; Mateo 1998) stellar system, located 86$\pm$5~kpc away from us 
   (Pietrzy{\'n}ski et al. 2008). The bulk of its stars are old ($>$10~Gyr old; 
   Da Costa 1984), with a small tail of stars at intermediate ages (6--10 Gyr; 
   Dolphin 2002; Tolstoy et al. 2003). A metallicity gradient is present in 
   Sculptor (Tolstoy et al. 2004), which is linked to an age gradient (de Boer 
   et al. 2011). Younger and more metal-rich stars concentrate towards the 
   centre. Under the hypothesis that Sculptor is not tidally disrupted, 
   Battaglia et al. (2008a) have estimated its total mass to be 
   $\mathscr{M}_{\mathrm tot}$($<$1.8 kpc) = (3.4$\pm$0.7$) \times $10$^{8}$ 
   M$_{\odot}$, making it more massive than previously thought.

   Recently, a very accurate SFH of Sculptor has been derived from deep, 
   wide-field CMDs covering a large fraction of the galaxy and going down to 
   the oldest main sequence turn-off (de Boer et al. 2011, 2012). The basic 
   features found in previous works have been confirmed: star formation took 
   place in Sculptor at early epochs and lasted several Gyr, from 14 to 7 Gyr 
   ago. During this period, the star formation proceeded at a steadily 
   decreasing rate (see de Boer et al. 2012, their figure~8). 

   The first studies of detailed chemical abundances in Sculptor are those of 
   Shetrone et al. (2003) and Geisler et al. (2005). They used the Ultraviolet 
   Echelle Spectrograph (UVES) on the Very Large Telescope (VLT) to determine 
   the abundances of several elements in 5 and 4 red giant branch (RGB) stars, 
   respectively. Notwithstanding the low number of stars probed, it was 
   immediately clear that the Sculptor dSph follows a chemical enrichment path 
   distinct from that of any of the Milky Way components. In the last few 
   years, thanks to the high multiplex capabilities of instruments such as 
   DEIMOS (Deep Imaging Multi-Object Spectrometer) on Keck~II and FLAMES (Fibre 
   Large Array Multi Element Spectrograph) on VLT, detailed abundances of 
   several elements have become available for hundreds of stars in Sculptor 
   (Kirby et al. 2009, 2010; North et al. 2012; Hill et al., in preparation, 
   see Tolstoy et al. 2009). Tight [X/Fe]--[Fe/H] relations have been found in 
   the metallicity range $-$2.5~$\le$ [Fe/H]~$\le -$1.0; in particular, the 
   [$\alpha$/Fe] ratio has been found to steadily decline with metallicity. For 
   a handful of extremely metal-poor stars below [Fe/H]~$\sim -$3.5 dex, a 
   detailed analysis of the abundances of the $\alpha$, iron-peak and heavy 
   elements has been carried out using high-resolution spectra taken with 
   Magellan/MIKE (Frebel et al. 2010) and VLT/UVES (Tafelmeyer et al. 2010). 
   Most recently, Starkenburg et al. (2013b) have enlarged the sample of 
   (extremely) low metallicity stars by adding seven objects with detailed 
   chemical abundances from spectra taken with X-shooter on the VLT.

   A well-defined stellar metallicity distribution function (MDF) for Sculptor 
   has been obtained from wide-field medium resolution Ca\,{\sevensize II} 
   triplet spectroscopy of more than 600 RGB stars (Battaglia et al. 2008b; 
   Starkenburg et al. 2010). Kirby et al. (2010) also present the MDF of 
   Sculptor's stars, for a smaller sample of $<$400 objects. The shape of the 
   MDFs from both studies is expected to be different due to the different 
   depth and radial extent of their data sets. There is good agreement for the 
   [Fe/H] values for stars in common between both datasets and the 
   high-resolution study of Hill et al. (in preparation). The MDF we use for 
   comparison with our model results in Sect.~\ref{sec:res} combines both data 
   sets in an attempt to mitigate the biases affecting individual studies (see 
   Appendix~\ref{sec:A} for details about the derivation of our more unbiased 
   MDF).

   The trend of the abundance ratios with [Fe/H] and the shape of the MDF 
   supply complementary information on the timescales of chemical enrichment in 
   Sculptor and provide an unprecedented bench mark for chemical evolution 
   studies. We will take advantage of the wealth of available data to sensibly 
   constrain the free parameters of our model.

   \section{The model}

   Starkenburg et al. (2013a) have studied the satellites of the Milky Way by 
   using a SAM of galaxy formation coupled to high-resolution \emph{N-}body 
   cosmological simulations. Their model galaxies match several observed 
   relations on the scale of the Milky Way and its satellites. The dwarf 
   galaxies display large variations in their SFHs, as observed. Based on the 
   SFHs, model satellites are identified that crudely resemble the Carina, 
   Sculptor and Fornax dSphs. The predicted MDFs, however, are too narrow with 
   respect to the observed ones (see Starkenburg et al. 2013a, their 
   figure~13). As discussed by the authors, this discrepancy is likely largely 
   due to the adoption of the instantaneous recycling approximation (IRA) in 
   their study and certainly deserves further analysis.

   In this paper, we compute the detailed chemical enrichment history of four 
   Sculptor-like dwarf galaxies identified from Starkenburg et al. (2013a; see 
   their figures~12 and 13 for SFHs and metallicity distributions of models 
   labelled Scl\,{\sevensize1}--{\sevensize 3} here, the fourth model
   labelled Scl\,{\sevensize4} was additionally selected for this work). The 
   Sculptor-like galaxies are selected on coarse luminosity ($-$11.8~$< M_V < 
   -$10.3) and dominant old stellar population criteria. In 
   Sections~\ref{sec:cos} and \ref{sec:sam} we concisely describe the adopted 
   cosmological framework and semi-analytic modelling. In Section~\ref{sec:gce} 
   we present the post-processing code that we have developed to compute the 
   evolution of chemical abundances in hierarchically growing systems. In 
   forthcoming papers of this series (Romano et al., in preparation; 
   Starkenburg et al., in preparation), we enlarge our sample and test the 
   predictions for satellite versus isolated objects. 

   \subsection{The cosmological simulations}
   \label{sec:cos}

   For the Aquarius project (Springel et al. 2008a) six halos were selected 
   from the lower resolution fully cosmological parent simulation, 
   Millennium~II, and resimulated at much higher resolution. The results in 
   this work were obtained from the level 2 Aquarius simulations with a 
   particle mass around $\sim$1~$\times$ 10$^{4}$~M$_{\odot}$. Tests on 
   different resolution runs, spanning a maximum range in particle mass of a 
   factor of 1800 in halo~A, show remarkably good convergence exceeding any 
   previous efforts (Springel et al. 2008a). Although the cosmological 
   parameters used for the simulation were based on the first-year results from 
   the Wilkinson Microwave Anisotropy Probe (WMAP) satellite and are no longer 
   consistent with the latest WMAP analysis (Komatsu et al. 2011), we do not 
   expect this to affect the results presented here significantly. Wang et 
   al. (2008), for instance, demonstrate that it is difficult to distinguish 
   between WMAP1 and WMAP3 in combination with SAMs for low-redshift galaxies. 
   Guo et al. (2013) make a similar comparison between WMAP1 and WMAP7. They 
   use the Millennium-II simulation, which allows for the formation of galaxies 
   down to $\sim$10$^8$~M$_{\odot}$, i.e. closer to the predicted mass of 
   Sculptor.

   The six main halos were selected to have dark matter masses in the range of 
   0.8--1.8~$\times$ $10^{12}$~M$_{\odot}$, consistent with the estimated mass 
   for the Milky Way (e.g. Wilkinson \& Evans 1999; Sakamoto et al. 2003; 
   Battaglia et al. 2005; Smith et al. 2007; Li \& White 2008; Xue et al. 2008; 
   Guo et al. 2010). The dark matter substructures that are the main interest 
   for our work, are identified using the code {\sevensize SUBFIND} (Springel 
   et al. 2001), which identifies self-bound structures within each 
   friends-of-friends group. A lower mass limit of 20 particles is used in 
   identifying the substructures. We refer the reader to Springel et al. 
   (2008a,b) for further information concerning the Aquarius simulations and to 
   Springel et al. (2001) for a description of {\sevensize SUBFIND}.

   \subsection{The semi-analytical model}
   \label{sec:sam}

   Substructure catalogues are used to construct merger history trees for all 
   self-bound halos and subhalos in the Aquarius simulations (Springel et al. 
   2005; De Lucia \& Blaizot 2007), which we are using as a backbone for 
   further modeling. In this semi-analytical approach we follow relevant 
   physical processes to study the flows and fate of baryons using relatively 
   simple expressions which are motivated and supported by observations 
   wherever possible. Our specific model applied for this work was described in 
   Starkenburg et al. (2013a) and stems from the models described in Kauffmann 
   et al. (1999), Springel et al. (2001) and De Lucia et al. (2004). Subsequent 
   updates have been made by Croton et al. (2006) and De Lucia \& Blaizot 
   (2007) and several minor adaptations were made to the model by De Lucia et 
   al. (2008), Li et al. (2009, 2010) and Starkenburg et al. (2013a) to model 
   more adequately the physics of galaxy formation and evolution within the 
   scale of a Milky Way environment.

\begin{figure}
   \psfig{figure=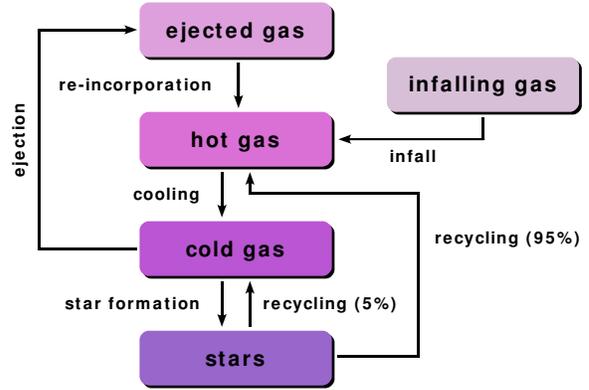,width=\columnwidth,angle=-90}
   \vspace{0.25cm}
      \caption{ A schematic chart of the flows of baryons in the adopted SAM. 
        The boxes represent the different phases of the baryons. The arrows 
        show how the baryons change phase, owing to the different processes 
        that affect them. Notice that in the \emph{`ejection model'} adopted in 
        this paper 95 per cent of newly produced metals are actually deposited 
        directly in the hot phase. Not shown here are the processes of 
        stripping. They can affect all phases and are implemented in the model 
        though.
              }
         \label{fig:chart}
   \end{figure}

   Physical processes modeled include reionization, cooling, star formation, 
   SN feedback and tidal stripping. A schematic diagram of several main 
   processes modeled and their interaction, is shown in Fig.~\ref{fig:chart}. 
   We refer the interested reader to Starkenburg et al. (2013a, and references 
   therein) for a detailed description of the semi-analytical prescriptions. 
   However, in order to enable a comparison with the numerical model for 
   chemical evolution used for the post-processing in this work (see 
   Section~\ref{sec:gce}), a review of the star formation and feedback 
   prescriptions is given below.

   The SFH is governed within the model by the amount of cold gas above a 
   critical density threshold within each system at each time,
   \begin{equation}
     \psi = \varepsilon {\mathscr M}_{\mathrm{sf}}/t_{\mathrm{dyn}},
   \end{equation}
   where $\varepsilon =$~0.03 represents the efficiency of the conversion of 
   gas into stars, ${\mathscr M}_{\mathrm{sf}}$ is the cold gas mass eligible 
   for star formation (which is assumed to form an exponential disk; see Mo et 
   al. 1998 for prescriptions) and $t_{\mathrm{dyn}} = 
   r_{\mathrm{disk}}/V_{\mathrm{vir}}$ is the dynamical time of the disk. 
   Following Kennicutt (1989), the critical density threshold for star 
   formation is
   \begin{equation} 
     \frac{\Sigma_{\mathrm{crit}}}{{\mathrm{M_\odot pc^{-2}}}} = 0.59 \, 
     \frac{V_{\mathrm{vir}}}{{\mathrm{km \, s^{-1}}}} \Big/ 
     \frac{r_{\mathrm{disk}}}{{\mathrm{kpc}}}. \\
     \label{eq:sigmacrit}
   \end{equation}
   A second, bursty mode of star formation is possible during minor or major 
   mergers, when (part of) the cold gas in the merging galaxies is turned into 
   stars. The model assumes that photo-dissociation from UV radiation will 
   prevent cooling from molecular hydrogen, thus does not allow gas to cool in 
   halos whose virial mass corresponds to a temperature below 10$^4$~K (the 
   atomic hydrogen cooling limit). This implies that the model does not take 
   into account a physical implementation of the first stars, which most 
   likely have to be cooled through the molecular hydrogen channel. A proper 
   implementation of first star physics would require assumptions on the 
   interplay between molecular hydrogen cooling and dissociation, as well as 
   the IMF for the first stars, which are not well understood.

   Within the SAM, IRA is adopted, meaning that immediate SN feedback is 
   modeled from the stars and that their finite ages are not taken into account 
   by the model. The recycled fraction, defined to be the ratio of the amount 
   of recycled gas to the total amount of gas that was initially converted into 
   stars, is 43 per cent in the SAM, using a Chabrier IMF. The feedback recipe 
   used is identical to the \emph{`ejection model'} described in Li et al. 
   (2009, 2010), where 95 per cent of the SN ejecta goes directly in the hot 
   phase, and includes a dependence on the halo potential well (i.e. 
   $\propto 1/V_{\mathrm{vir}}^{2}$) to determine the amount of cold gas 
   affected by the energy output from SNe. The fraction of SN ejecta that is 
   instantaneously returned to the ISM (5 per cent) is mixed with the cold gas 
   before reheating. The reheated cold material is stored in a separated 
   component of ejected gas and can be reincorporated into the hot gas 
   reservoir available at later times. In the majority of this work, we assume 
   reheated cold gas never returns to the ISM. However, we also test the effect 
   of allowing some of it to return at later times (see Sect.~\ref{sec:rec}).


   \subsection{The galactic chemical evolution tool}
   \label{sec:gce}

   The information as schematically represented in Fig.~\ref{fig:chart} is 
   subsequently post-processed to determine the chemical evolution of the 
   galaxy. We use the full history for the galaxy, including possible merging 
   events. For each baryonic state described in Fig.~\ref{fig:chart}, the 
   information from all progenitors is added together. We note, however, that 
   late (major) merging events are rare in all our Sculptor-type galaxies that 
   typically become a satellite of the Milky Way-like galaxy quite early in 
   their life. The adopted SAM additionally does not provide spatially resolved 
   information about the baryonic content within the targeted galaxies. 
   Therefore, we use a one zone approximation and follow the evolution of 
   tracked elements in the cold gas out of which stars form irrespective of 
   which progenitor they formed in or the distance from the galaxy centre, by 
   means of the following equations:
   \begin{equation}
     \frac{{\mathrm d}{\mathscr M}_{i}(t)}{{\mathrm d}t} = -X_{i}(t)\psi(t) 
     + (1-F) {\mathscr R}_{i}(t)+\frac{{\mathrm d}{\mathscr M}^{\mathrm{in}}_{i}
       (t)}{{\mathrm d}t}-\frac{{\mathrm d}{\mathscr M}^{\mathrm{out}}_{i}
       (t)}{{\mathrm d}t},
     \label{eq:basic}
   \end{equation}
   where ${\mathscr M}_{i}(t) = X_{i}(t){\mathscr M}_{\mathrm{cold}}(t)$ is the 
   cold gas mass in the form of the element $i$ at the time $t$, $X_{i}(t)$ is 
   the abundance by mass of the element $i$ in the cold gas at the time $t$ and
   the summation of all $X_{i}(t)$ is equal to unity. $\psi(t)$ is the star 
   formation rate (SFR), ${\mathscr R}_{i}(t)$ is the production rate of the 
   element $i$ by dying stars. $F$ is the fraction of the stellar ejecta that 
   goes directly into the hot gas phase (see previous section and Li et al. 
   2010). The last two terms on the right-hand side of 
   equation~(\ref{eq:basic}) account for the addition/loss of element $i$ 
   to/from the cold gas phase owing to: (i) cooling of hot gas, (ii) 
   interactions with other galaxies\footnote{Both our SAM and chemical 
     post-processing code have in principle the ability to deal with tidal 
     stripping of gas, as well as stars. In the specific models studied such 
     events do not occur however.} and (iii) ejection of gas heated by SN 
   explosions. The main physical processes entering the right-hand side of 
   equation~(\ref{eq:basic}) are sketched in Fig.~\ref{fig:chart}.

   \subsubsection{Stellar nucleosynthesis}
   \label{sec:nuc}

   In our model, the production rate of the element $i$ at the time $t$ is 
   computed by taking into account in detail the contributions of stars of 
   different initial masses (lifetimes) and chemical compositions (using the 
   {\it Q}-matrix formalism; Talbot \& Arnett 1973), as well as the presence of 
   binary systems ending up as SNeIa:
   \begin{equation}
     {\mathscr R}_{i}(t) = {\mathscr R}_{i}^{\mathrm{LIMS}}(t) + 
     {\mathscr R}_{i}^{\mathrm{SNII}}(t) + {\mathscr R}_{i}^{\mathrm{SNIa}}(t).
   \end{equation}
   ${\mathscr R}_{i}^{\mathrm{LIMS}}(t)$, ${\mathscr R}_{i}^{\mathrm{SNII}}(t)$ 
   and ${\mathscr R}_{i}^{\mathrm{SNIa}}(t)$ are the rates at which low- and 
   intermediate-mass stars, core-collapse SNe and SNeIa, respectively, restore 
   each element to the ISM. For low- and intermediate-mass stars, it reads
   \begin{equation}
     {\mathscr R}_{i}^{\mathrm{LIMS}}(t) = (1-{\mathscr B}) 
     \int_{m_{l}(t)}^{m_{\mathrm{WD}}} \varphi(m) 
     \psi(t - \tau_{m}){\mathscr{Q}}_{mi}(t - \tau_{m}) {\mathrm{d}}m,
   \end{equation}
   where $m_{l}(t)$ is the turn-off mass at the time $t$, $m_{\mathrm{WD}} =$ 
   8~M$_\odot$ is the maximum mass for white dwarf formation, $\tau_{m}$ is the 
   lifetime of the star of initial mass $m$, $\varphi(m)$ is the IMF. Here we 
   interpolate linearly between tabulated stellar lifetimes from Schaller et 
   al. (1992) and assume an extrapolation of the Salpeter (1955) IMF in the 
   mass range 0.1--100~M$_\odot$, unless otherwise stated. Following Maeder 
   (1992), the quantities ${\mathscr{Q}}_{mi}(t - \tau_{m})$ are the summation 
   of two terms,
   \begin{equation}
     {\mathscr{Q}}_{mi}(t - \tau_{m}) = X_{i}(t - \tau_{m}) m_{\mathrm{ej}}(m) 
     + mp_{i}(m),
     \label{eq:lims}
   \end{equation}
   that consistently compute the mass ejected in the form of the element $i$ 
   that was already present at the stellar birth ---first term on the 
   right-hand side of equation~(\ref{eq:lims})--- and the newly synthesized one 
   according to specific yields tables ---second term on the right-hand side of 
   equation~(\ref{eq:lims}); $m_{\mathrm{ej}}(m)$ and $p_{i}(m)$ are, 
   respectively, the total mass ejected by a star of mass $m$ during its 
   lifetime and the stellar yield (Tinsley 1980). The quantity ${\mathscr B}$ 
   is the realization probability for SNeIa. It is a free parameter of the 
   model and is different from zero only in a restricted mass range 
   (3--16~M$_\odot$; see next paragraph). Here ${\mathscr B} =$ 0.03, i.e. we 
   set ${\mathscr B}$ to the same value which allows us to reproduce the 
   present-day SNIa rate in the disk of the Milky Way (Li et al. 2011). 
   Similarly, for massive stars,
   \begin{equation}
     {\mathscr R}_{i}^{\mathrm{SNII}}(t) = (1-{\mathscr B}) \int_{m_{cc}}^{m_{u}} 
     \varphi(m) \psi(t - \tau_{m}){\mathscr{Q}}_{mi}(t - \tau_{m}) {\mathrm{d}}m,
   \end{equation}
   where $m_{u} =$ 100~M$_\odot$ is the upper mass limit of the IMF and $m_{cc} 
   =$ 8~M$_\odot$ is the minimum mass for core-collapse SNe.

   The rate at which SNeIa restore their nucleosynthetic products to the ISM is 
   a strong function of the adopted SFR and progenitor model and, to a lesser 
   extent, of the adopted stellar lifetimes and IMF (Greggio \& Renzini 1983; 
   Matteucci \& Greggio 1986; Kobayashi et al. 1998; Matteucci \& Recchi 2001; 
   Greggio 2005; Matteucci et al. 2006, 2009; Kobayashi \& Nomoto 2009). 
   Following Greggio \& Renzini (1983) and Matteucci \& Greggio (1986), it can 
   be written
   \begin{eqnarray}
     \lefteqn{ {\mathscr R}_{i}^{\mathrm{SNIa}}(t) = {\mathscr B} 
     \int_{m_{\mathrm{b}_{min}}(t)}^{m_{\mathrm{b}_{max}}(t)} \varphi(m_{\mathrm b}) 
     \int_{\mu_{\mathrm{min}}}^{\mu_{\mathrm{max}}} f(\mu)} 
     \nonumber \\
     & & {} \psi(t - \tau_{m_2}) {\mathscr{Q}}_{m_1i}(t - \tau_{m_2}) 
     {\mathrm{d}}\mu \, {\mathrm{d}}m_{\mathrm b},
   \end{eqnarray}
   where $m_{\mathrm b} = m_1 + m_2$ is the total mass of the binary system and 
   $\mu = m_2/m_{\mathrm b}$. The masses $m_{\mathrm{b}_{min}} (t)$ and 
   $m_{\mathrm{b}_{max}}(t)$ are the minimum and maximum masses, respectively, 
   of the contributing systems at the time $t$; the minimum and maximum values 
   that they can take are 3~M$_\odot$ and 16~M$_\odot$, respectively. $f(\mu)$ 
   is the distribution function for the mass of the secondary star with respect 
   to the total mass of the system and is taken from Greggio \& Renzini (1983), 
   with $\mu_{\mathrm{min}} =$ max\big($m_2/m_{\mathrm b}$, ($m_{\mathrm b} - 
   8$)/$m_{\mathrm b}$\big), $\mu_{\mathrm{max}} =$ 0.5. The lifetime of the 
   secondary star, $\tau_{m_2}$, is the clock for the explosion. We refer to 
   Matteucci \& Recchi (2001, and references therein) for more details. We 
   remark that no metallicity effect leading to the inhibition of SNeIa at low 
   metallicity (Kobayashi et al. 1998; Kobayashi \& Nomoto 2009) is included in 
   the present work. We will analyze the effect of adopting different 
   prescriptions for SNeIa on our model results in a forthcoming paper.

   As for single stars, in this work we adopt the metallicity-dependent yields 
   from van den Hoek \& Groenewegen (1997) for LIMS and Woosley \& Weaver 
   (1995) for massive stars. Uncertainties in the iron yields are a factor of 
   two; following Timmes et al. (1995), we halve the yields of iron from 
   massive stars in the original Woosley \& Weaver's tables (see also Goswami 
   \& Prantzos 2000; Romano et al. 2010a). As for binary systems giving rise to 
   SNIa events, we use Iwamoto et al.'s (1999) yields (their model W7) apart 
   from Mn, for which we use the prescriptions of Cescutti et al. (2008; see 
   also Romano et al. 2011). It is worth emphasizing that this combination of 
   stellar yields ensures a good fit to the [X/Fe]--[Fe/H] relations of most 
   chemicals in the solar neighbourhood, at metallicities typical of dSphs 
   (Romano et al. 2010a, their figure~22, model~1). In general, our code is 
   structured in such a way that it can be easily supplied with different yield 
   sets.

   \subsubsection{Gas flows and star formation: the cosmological context}
   \label{sec:gas}

   In the classical approach (see next section), the dwarf galaxy is treated as 
   an isolated object, with a smooth accretion of infalling gas. For the models 
   computed in a cosmological context, we adopt the SFH and gas flows from the 
   cosmological simulations and SAM described in Sects.~\ref{sec:cos} and 
   \ref{sec:sam}. It reads:
   \begin{equation}
     \frac{{\mathrm d}{\mathscr M}^{\mathrm{in}}_{i}(t)}{{\mathrm d}t} = 
     X^{\mathrm{cool}}_{i}(t) \frac{{\mathrm d}{\mathscr 
         M}_{\mathrm{cool}}(t)}{{\mathrm d}t},
   \end{equation}
   where $X^{\mathrm{cool}}_{i}(t) = X^{\mathrm{hot}}_{i}(t)$ is the abundance by 
   mass of the element $i$ in the cooling flow at the time $t$ and ${\mathscr 
     M}_{\mathrm{cool}}(t)$ is the total mass accreted from the hot gas 
   reservoir at the time $t$, and 
   \begin{equation}
     \frac{{\mathrm d}{\mathscr M}^{\mathrm{out}}_{i}(t)}{{\mathrm d}t} = 
     X_{i}(t) \xi_{i} \frac{{\mathrm d}{\mathscr 
         M}_{\mathrm{reheat}}(t)}{{\mathrm d}t},
     \label{eq:xi}
   \end{equation}
   where $X_{i}(t)$ is the abundance by mass of the element $i$ in the ISM at 
   the time $t$ and ${\mathscr M}_{\mathrm{reheat}}(t)$ is the ISM mass heated 
   by SN explosions and ejected at the time $t$. $\xi_{i}$ can be set to 1 for 
   all elements (as in Li et al.'s 2010 paper) or to higher values for metals, 
   to mimic a differential outflow. 


   Since the hypothesis of IRA is relaxed in our computations, our 
   ${\mathscr M}_{\mathrm{cold}}(t)$ value may differ from the corresponding SAM 
   one (the differences are within a few per cent). Therefore, at each time 
   step we readjust it to its SAM value. We do this after the gas mixture is 
   assigned the proper chemical composition. It is worth stressing again here 
   that the SAM may provide more than one value for 
   ${\mathscr M}_{\mathrm{cold}}(t)$ at each time step, depending on the number 
   of progenitors. In order to deal with this, at each time step we sum up all 
   the values of ${\mathscr M}_{\mathrm{cold}}(t)$ from the SAM. The vast 
   majority of such merging events occur in the early Universe in our models.

   The inclusion of SNeIa and their delay times in the model has a huge impact 
   on the predicted MDF and [X/Fe] versus [Fe/H] behaviour, as we will see in 
   the next section.

   \subsubsection{Gas flows and star formation: the classical approach}
   \label{sec:classic}

   For the purpose of comparison with the models computed within a cosmological 
   framework, we also run a classical model for Sculptor. In the frame of such 
   a model, a simple Schmidt (1963) star formation law is assumed, as usually 
   done in classical chemical evolution studies:
   \begin{equation}
     \psi(t) = \nu {\mathscr M}^{k}_{\mathrm{cold}}(t),
   \end{equation}
   where $\nu$ and $k$ are free parameters of the model, that are adjusted to 
   reproduce the observations. The $\nu$ parameter is the star formation 
   efficiency. It is related to $\varepsilon$, its counterpart in the SAM, 
   through the relation $\nu$~= $\varepsilon/t_{\mathrm{dyn}}$, where 
   $t_{\mathrm{dyn}}$ is the dynamical time and $\varepsilon$~= 0.03 in the 
   adopted SAM (see Sect.~\ref{sec:sam}). Hence, the two quantities must not be 
   directly compared. We do not consider a threshold gas density for star 
   formation in the classic model. The raw material for star formation is 
   accreted according to a time-decaying infall rate:
   \begin{equation}
     \frac{{\mathrm d}{\mathscr M}^{\mathrm{in}}_{i}(t)}{{\mathrm d}t} = 
     X^{\mathrm{in}}_{i}(t) A {\mathrm e}^{-t/\tau},
     \label{eq:in}
   \end{equation}
   with the accreted matter being usually assigned a primordial chemical 
   composition, $X^{\mathrm{in}}_{i}(t) = X^{0}_{i}$. The normalization constant, 
   $A$, obeys the boundary condition 
   \begin{equation}
     \int_{0}^{t_{\mathrm{now}}} A {\mathrm e}^{-t/\tau} {\mathrm d}t = {\mathscr 
       M}_{\mathrm{acc}},
   \end{equation}
   where ${\mathscr M}_{\mathrm{acc}}$ is the total mass ever accreted by the 
   system; $t_{\mathrm{now}}$ and $\tau$ are the age of the Universe and the 
   infall time scale, respectively. We adopt $t_{\mathrm{now}} =$ 13.8 Gyr (cf. 
   Bennett et al. 2012) for the present-day age of the Universe. SNe of all 
   types inject energy in the surrounding ISM. When the thermal energy of the 
   gas heated by SN explosions exceeds its binding energy, a galactic wind 
   eventually develops. The rate of gas loss via galactic wind is 
   \begin{equation}
     \frac{{\mathrm d}{\mathscr M}^{\mathrm{out}}_{i}(t)}{{\mathrm d}t} = 
     X^{\mathrm{out}}_{i}(t) w_{i} {\mathscr M}_{\mathrm{cold}}(t),
     \label{eq:out}
   \end{equation}
   where $X^{\mathrm{out}}_{i}(t) = X_{i}(t)$, namely, the abundance of each 
   element in the wind is the same as in the ISM. The $w_{i}$ terms are further 
   parameters of the model that describe the efficiency of the outflow for each 
   element; they are set all to the same value in the case of normal wind, but 
   can take a higher value for metals in the case of differential, 
   metal-enriched winds (Mac Low \& Ferrara 1999; Recchi et al. 2001; Fujita et 
   al. 2004).

   In classical models, there is no distinction between cold and hot gas 
   phases. The freshly produced metals are assumed to cool down in short time 
   scales (i.e., shorter than the computation time steps) and $F =$ 0 in 
   equation~(\ref{eq:basic}). This means to have instantaneous mixing and 
   instantaneous cooling of newly released metals in the models.

   We set $\nu$~= 0.02 Gyr$^{-1}$, $k$~= 1, ${\mathscr M}_{\mathrm{acc}}$~= 
   1.7~$\times$~10$^{8}$ M$_{\odot}$, $\tau$~= 5~$\times$ 10$^7$ yr and 
   $w_{\mathrm{H, He}}$~= 0.5 Gyr$^{-1}$, $w_{\mathrm{metals}} \simeq$ 1 
   Gyr$^{-1}$. This is model Scl\,{\sevensize T}. The parameters of this model 
   have been tuned to reproduce, as best as we can within this simple working 
   framework: (i) the CMDs (see de Boer 2012) and the detailed SFH of Sculptor 
   inferred from observations (de Boer et al. 2012); (ii) the trends of several 
   abundance ratios with [Fe/H] in Sculptor's stars; (iii) the MDF 
   representative of the full Sculptor stellar population (see 
   Appendix~\ref{sec:A}) and (iv) the (likely) absence of neutral hydrogen in 
   the galaxy at the present time (Grcevich \& Putman 2009).

   \section{Results}
   \label{sec:res}


   \begin{table}
     \caption{ Properties of the `cosmological' Sculptor galaxy models.}
     \begin{tabular}{lcccc}
       \hline
       Model & $\mathscr{M}_{\mathrm{stars}}$ & $\mathscr{M}_{\mathrm{cold\,gas}}$ 
                 & $\langle$[Fe/H]$\rangle_{\mathrm{stars}}$ & $M_V$ \\
             & (10$^6$ M$_\odot$) & (10$^6$ M$_\odot$) & (dex) & (mag) \\
       (1) & (2) & (3) & (4) & (5) \\
       \hline
       Scl\,{\sevensize 1} & 6.1 & 2.2 & $-$0.9 ($-$1.6$^a$) & $-$10.9 \\
       Scl\,{\sevensize 2} & 2.7 & 14. & $-$1.7 & $-$10.3 \\
       Scl\,{\sevensize 3} & 4.3 & 0.1 & $-$1.1 ($-$1.5$^a$) & $-$10.7 \\
       Scl\,{\sevensize 4} & 13. & 7.2 & $-$1.4 & $-$11.8 \\
       \hline
       Observed$^b$ & $\sim$8.0 & 0.234 & $-$1.9 & $-$11.1 \\
       \hline
       \end{tabular}
       \label{tab:mod}

       \medskip
       \emph{Note.} Different columns list: (1) the model name; (2) the 
       present-day stellar mass; (3) the present-day gaseous mass; (4) the mean 
       metallicity of the stellar populations; (5) the $V$-band absolute 
       magnitude. Observed values are given in the last row.\\
       $^{a}$With last star formation burst excluded.\\
       $^{b}$ The stellar mass is estimated from the SFH of de Boer et al. 
       (2012). The cold gas mass is the neutral hydrogen mass; notice the 
       detection is ambiguous due to the numerous intervening clouds that could 
       be mistaken for gas associated with Sculptor (Grcevich \& Putman 2009). 
       The mean stellar metallicity is computed from values in 
       Table~\ref{tab:appx} of this work. The $V$-band absolute magnitude is 
       obtained from the apparent magnitude reported by Irwin \& Hatzidimitriou 
       (1995), by assuming a distance of 86 kpc for Sculptor.
   \end{table}



   \begin{figure*}
     \begin{tabular}{cc}
       \psfig{file=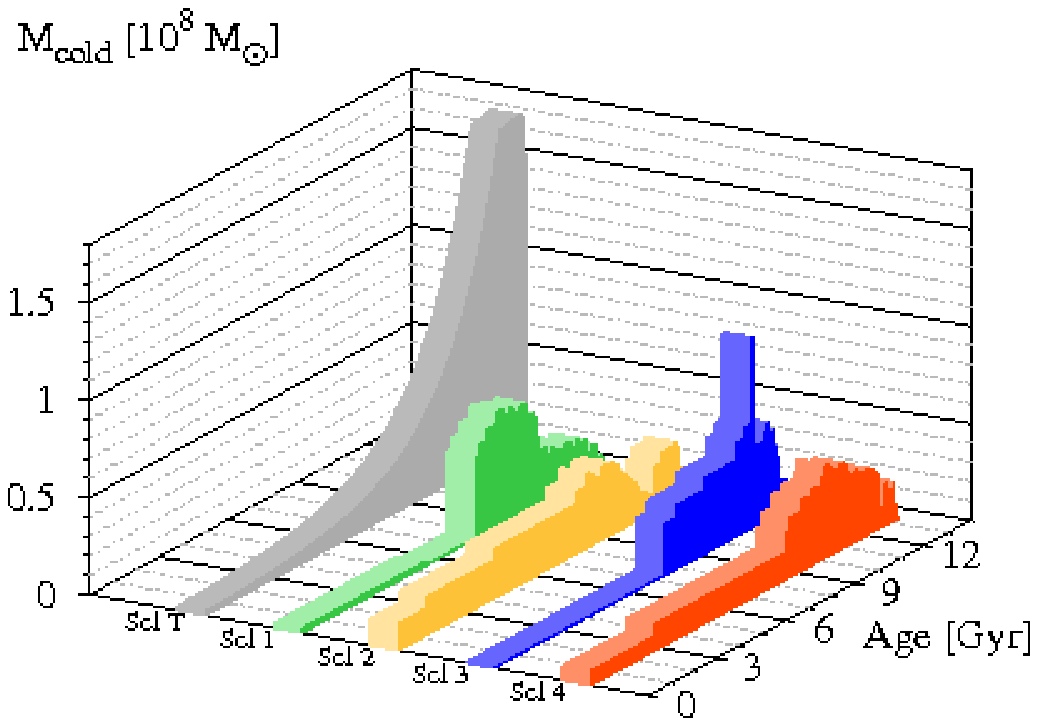,width=0.5\linewidth,clip=} &
       \psfig{file=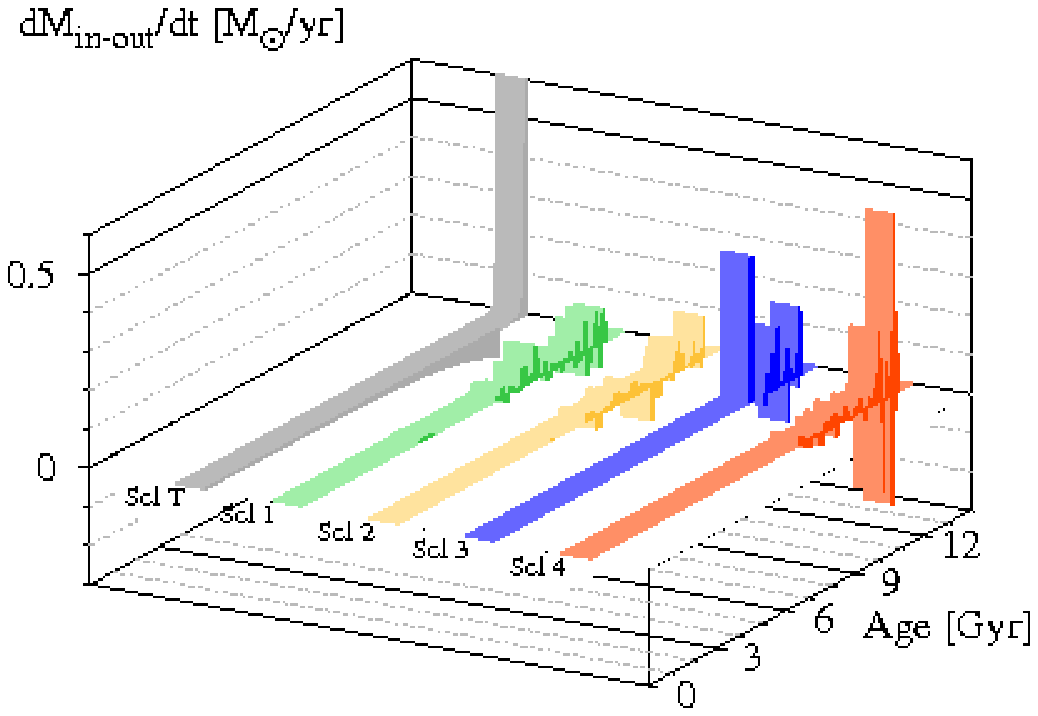,width=0.5\linewidth,clip=} \\
       \psfig{file=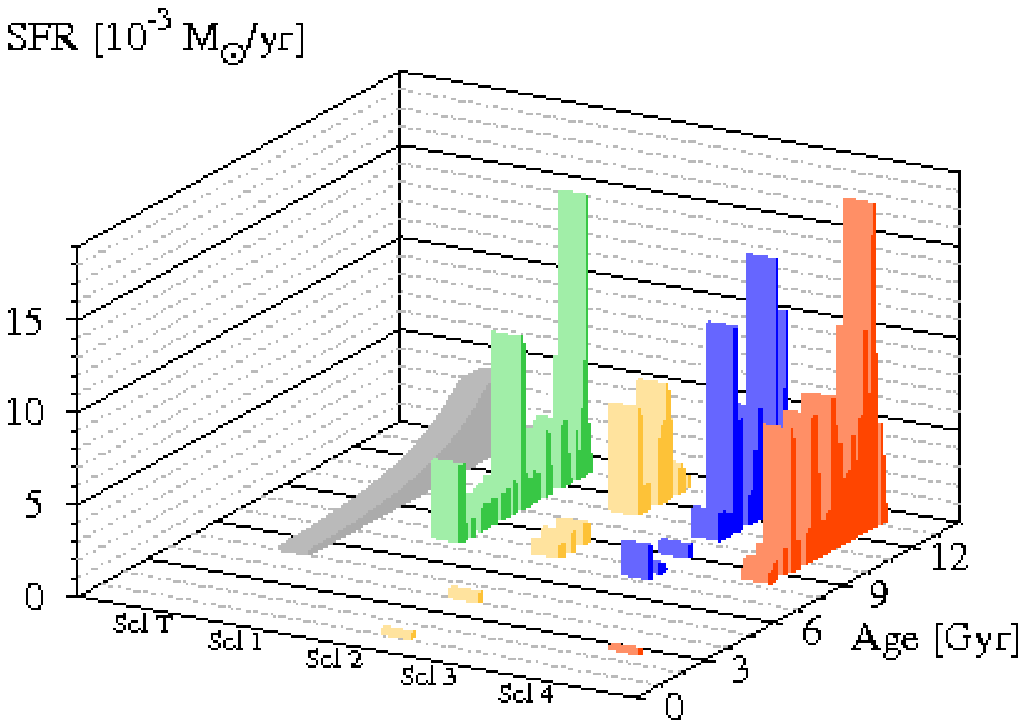,width=0.5\linewidth,clip=} &
       \psfig{file=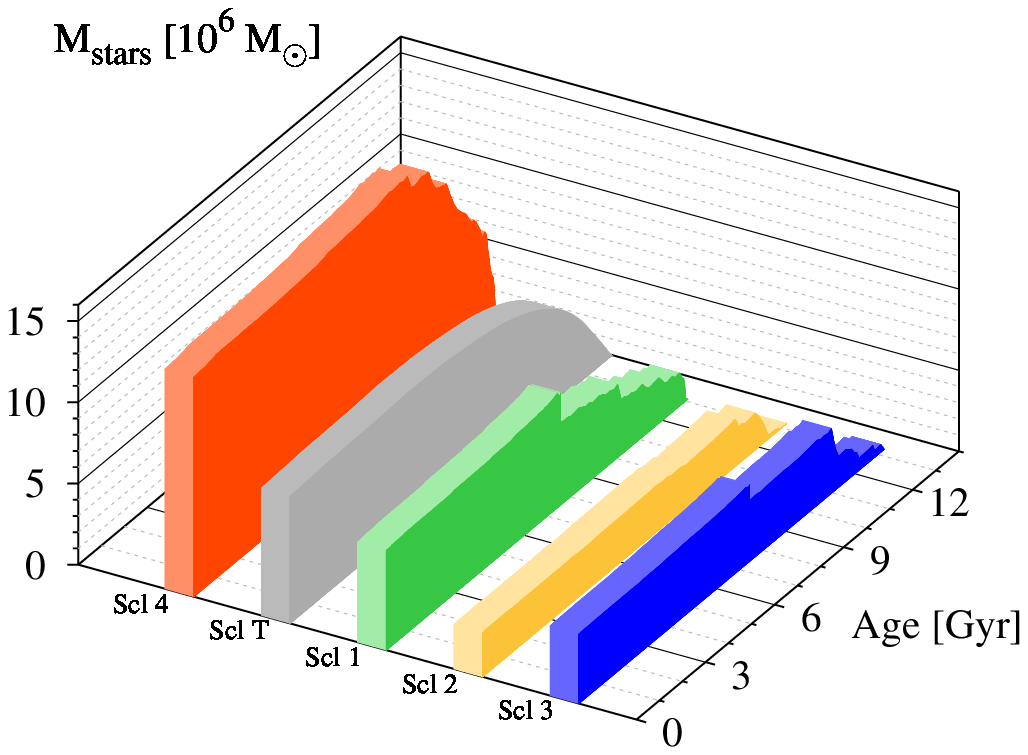,width=0.5\linewidth,clip=}
     \end{tabular}
      \caption{ Counterclockwise from top left: evolution of the cold gas 
        masses, SFRs and cumulative stellar masses for models Scl\,{\sevensize 
          T} (grey walls), Scl\,{\sevensize 1} (green walls), Scl\,{\sevensize 
          2} (yellow walls), Scl\,{\sevensize 3} (blue walls) and 
        Scl\,{\sevensize 4} (red walls). In the top right box we show the 
        difference between incoming and outcoming cold gas fluxes for the same 
        models.
              }
         \label{fig:summary}
   \end{figure*}


   In this section we present our results concerning the detailed chemical 
   composition of four Sculptor candidates selected by Starkenburg et al. 
   (2013a) on luminosity ($-$11.8 $< M_V < -$10.3) and dominant old stellar 
   population criteria. The outputs of these `cosmologically motivated' models 
   are compared to the ones from a classical model, as well as to the relevant 
   data. All the theoretical abundance ratios discussed in this work are 
   normalized to the solar abundances by Grevesse \& Sauval (1998).

   In Fig.~\ref{fig:summary} we show the evolution of the cold gas masses, the 
   net change in cold gas masses, the SFRs and the cumulative stellar masses 
   for the four models listed in Table~\ref{tab:mod}. The cold gas masses, gas 
   flows and SFRs are the input of the post-processing code described in the 
   previous section. A classical model of chemical evolution meant to meet the 
   main observational constraints for Sculptor (see Sect.~\ref{sec:classic}) is 
   considered as well (Fig.~\ref{fig:summary}, grey walls).

   It is immediately seen (Fig.~\ref{fig:summary}, top left box) that, at early 
   evolutionary stages, a gross cold gas amount, of more than 1.5~$\times$ 
   10$^8$~M$_\odot$, characterizes the classical model, which leaves behind a 
   small stellar system of only ${\mathscr M}_{\mathrm{stars}}$~= 7.8~$\times$ 
   10$^6$ M$_\odot$ (Fig.~\ref{fig:summary}, bottom right box). In contrast, in 
   hierarchically growing systems a much lower (up to one order of magnitude) 
   cold gas mass is predicted at early epochs. Yet, the current stellar masses 
   are in between 2.7 and 13~$\times$ 10$^6$ M$_\odot$, reflecting a more 
   efficient star formation in these models. Another striking difference 
   regards the history of mass assembly. While in the classical approach it is 
   fairly simple ---a short phase of strong gas accretion, followed by a much 
   longer period in which mass loss dominates--- the models computed within the 
   hierarchical scheme of galaxy formation display far more complex patterns 
   (see Fig.~\ref{fig:summary}, top right box). Because of the assumed critical 
   density threshold for star formation, the SFR of the cosmological models may 
   be zero even if the gas content is different from zero 
   (Fig.~\ref{fig:summary}, bottom left box). Model~Scl\,{\sevensize 2} 
   predicts a present-day cold gas mass significantly higher than the limit on 
   the neutral hydrogen mass suggested by Grcevich \& Putman (2009) for the 
   Sculptor dSph (Table~\ref{tab:mod}). Therefore, it can be ruled out as a 
   good Sculptor replica. Also Model~Scl\,{\sevensize 4} predicts a present-day 
   cold gas mass higher than observed.

   It is worthwhile mentioning here that, while the adopted SAM reproduces the 
   luminosity function of Milky Way's satellites well, the predicted stellar 
   mass versus dark matter halo mass relation is offset with respect to the 
   extrapolations of the Guo et al. (2010) and Moster et al. (2010) abundance 
   matching relations (but it is in accordance with hydrodynamical simulations; 
   see Starkenburg et al. 2013a).

   \subsection{Age-metallicity relation and metallicity distribution 
     function}
   \label{sec:amrmdf}

\begin{figure}
   \psfig{figure=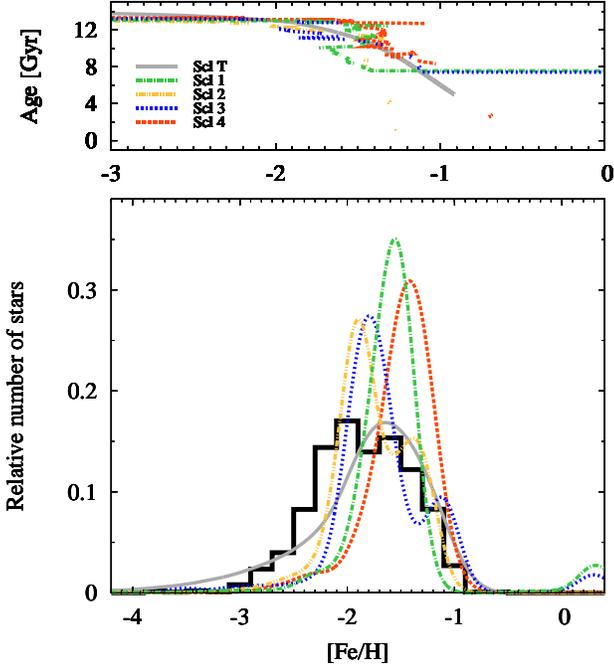,width=\columnwidth}
      \caption{ Age-metallicity relations (upper panel) and MDFs (bottom panel) 
        for our model galaxies Scl\,{\sevensize 1} (green dot-dashed lines), 
        Scl\,{\sevensize 2} (yellow dot-dot-dashed lines), Scl\,{\sevensize 3} 
        (blue dotted lines), Scl\,{\sevensize 4} (red dashed lines) and 
        Scl\,{\sevensize T} (grey solid lines). The theoretical MDFs have been 
        smoothed by a Gaussian function with a variance equal to the data 
        error, 0.15 dex. The black solid histogram is the observational MDF we 
        obtain by combining the DART CaT sample (Battaglia et al. 2008b; 
        Starkenburg et al. 2010) with the sample by Kirby et al. (2010), as 
        specified in Appendix~\ref{sec:A}.
              }
         \label{fig:mdf}
   \end{figure}

   At early times, because of the lower amounts of diluting gas the models 
   computed within the hierarchical picture for structure formation typically 
   reach higher metallicities than the classical one. This is clearly seen in 
   Fig.~\ref{fig:mdf}, upper panel, which displays the age-metallicity 
   relations of all our model galaxies. Model~Scl\,{\sevensize 2} is a notable 
   exception, in that it has the smoothest metallicity increase during the 
   first Gyr of evolution, due to its low-level star formation at those old 
   ages. The abrupt rise in metallicity characterizing models~Scl\,{\sevensize 
     1} and Scl\,{\sevensize 3} at an age of $\sim$7.5~Gyr (the horizontal 
   portions of the green dot-dashed and blue dotted lines, respectively, on the 
   right in Fig.~\ref{fig:mdf}, upper panel), is due to the last, strong bursts 
   of star formation, that almost exhaust the cold gas in these model galaxies 
   (Fig.~\ref{fig:summary}, green and blue walls, bottom and top left boxes).

   In Fig.~\ref{fig:mdf}, upper panel (as well as in all the following figures 
   of this paper), some gaps appear in the theoretical curves for 
   models~Scl\,{\sevensize 1}, Scl\,{\sevensize 2}, Scl\,{\sevensize 3} and 
   Scl\,{\sevensize 4}. They arise because we do not plot the portions of the 
   curves that correspond to halts in star formation (the curves reflect the 
   chemical composition of the ISM at each time; if the SFR at that time is 
   zero, no stars form with the corresponding chemical composition). For the 
   same reason, the curve relative to model~Scl\,{\sevensize T} ends at an age 
   of 5~Gyr, i.e. when the star formation is stopped in this model. We impose a 
   cut-off time for star formation in model~Scl\,{\sevensize T} to better match 
   the SFH of Sculptor derived by de Boer et al. (2012). However, it is worth 
   emphasizing that the star formation activity in model~Scl\,{\sevensize T} is 
   fading at ages less than 5 Gyr and would die out anyway because of the 
   strong mass loss through the outflow and scarce replenishment of gas from 
   outside.


   \begin{figure}
   \psfig{figure=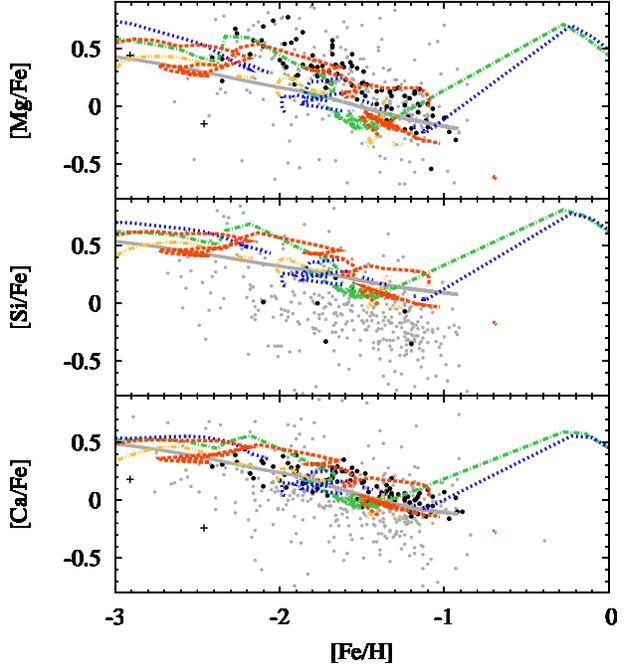,width=\columnwidth}
      \caption{ [X/Fe] versus [Fe/H] for several $\alpha$-elements (Mg, Si and 
        Ca) predicted by models Scl\,{\sevensize 1} (green dot-dashed lines), 
        Scl\,{\sevensize 2} (yellow dot-dot-dashed lines), Scl\,{\sevensize 3} 
        (blue dotted lines), Scl\,{\sevensize 4} (red dashed lines) and 
        Scl\,{\sevensize T} (grey solid lines). The grey dots represent the 
        data from medium-resolution spectra by Kirby et al. (2009), the black 
        dots those from high-resolution spectra by Shetrone et al. (2003, 5 
        stars), Geisler et al. (2005, 4 stars) and Hill et al. (in preparation, 
        89 stars; see Tolstoy et al. 2009). Data from Starkenburg et al. 
        (2013b) are shown as crosses.
              }
         \label{fig:mgsica}
   \end{figure}


   An important diagnostic to test our models is the MDF of long-lived stars, 
   which allows us to fine-tune some important parameters regarding chemical 
   evolution (see Sect.~\ref{sec:unc} for a thorough discussion). In 
   Fig.~\ref{fig:mdf}, bottom panel, our theoretical MDFs (lines) are compared 
   to the observational one (solid histogram; see Appendix~\ref{sec:A}). The 
   theoretical MDFs have been convolved with a Gaussian ($\sigma$~= 0.15) to 
   take the observational errors into account. Model~Scl\,{\sevensize T} 
   successfully reproduces the existence of stars with $-$4~$<$ [Fe/H]~$< -$2.5 
   in the right percentage (by construction), but the number of very metal-poor 
   stars in the metallicity range $-$2.5~$<$ [Fe/H]~$< -$2 is underestimated by 
   about 25 per cent. The `cosmological' models severely underestimate the 
   number of very metal-poor stars\footnote{We note that, while in the adopted 
     SAM many Sculptor analogs are found when looking at the SFH ---most model 
     satellites are dominated by stars formed at old ages--- candidates of 
     comparable luminosity tend to have slightly higher average metallicity 
     than Sculptor (Starkenburg et al. 2013a). This is in qualitative agreement 
     with observations of dwarf galaxies, using the luminosity and metallicity 
     as quoted in this paper the Sculptor dSph is placed on the lower edge of 
     the observed scatter in the luminosity-metallicity relation.}. 
   Models~Scl\,{\sevensize 2} and Scl\,{\sevensize 3} show a prominent peak at 
   [Fe/H]~$\sim -$1.9 dex, in fairly good agreement with the observations, but 
   predict unwanted secondary peaks at higher metallicities as well. The last, 
   intense bursts of star formation closing the SFHs of models~Scl\,{\sevensize 
     1} and Scl\,{\sevensize 3} produce small fractions of stars (about 5 per 
   cent of the total) with supersolar metallicities, that have no match in the 
   real galaxy. According to models Scl\,{\sevensize 1}, Scl\,{\sevensize 2}, 
   Scl\,{\sevensize 3} and Scl\,{\sevensize 4}, basically no stars are expected 
   below [Fe/H]~$\simeq -$2.5 dex. In this context, a detailed implementation 
   of the formation of the first stars would be an interesting addition to the 
   current SAM. Implementation of those physical processes could potentially 
   alter the first phases of star formation in the dwarf galaxies and, hence, 
   alter the MDF at the very metal-poor end. We also note that a lowering of 
   the initial mass limit for black hole formation in stellar populations of 
   very low-metallicity would result in a diminished production of iron during 
   early galactic evolution (more heavy elements would be swallowed by black 
   holes in this case; cf. Maeder 1992; Woosley \& Weaver 1995), thus leading 
   to a better fit of the very low-metallicity tail of the MDF.

   \subsection{Abundance ratios}
   \label{sec:ratios}


   \begin{figure}
   \psfig{figure=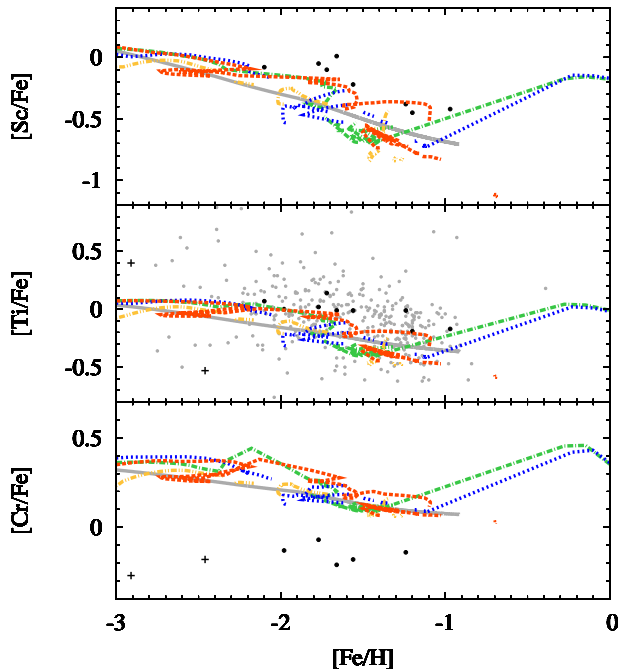,width=\columnwidth}
      \caption{ Same as Fig.~\ref{fig:mgsica}, for Sc, Ti and Cr.
              }
         \label{fig:scticr}
   \end{figure}



   \begin{figure}
   \psfig{figure=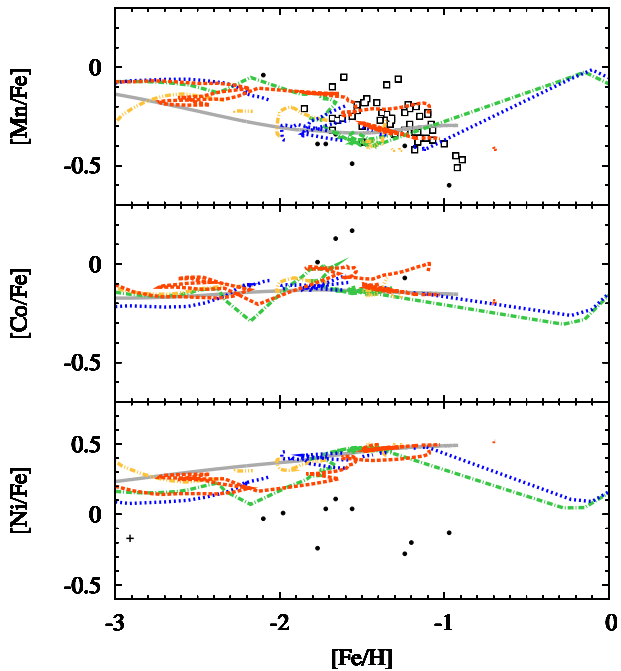,width=\columnwidth}
      \caption{ Same as Fig.~\ref{fig:mgsica}, for Mn, Co and Ni. Open squares 
        (top panel) are data from North et al. (2012).
              }
         \label{fig:mnconi}
   \end{figure}


   In Figs.~\ref{fig:mgsica} to \ref{fig:mnconi} we show our model predictions 
   on the [X/Fe] versus [Fe/H] behaviour for several $\alpha$ and iron-peak 
   elements in the stars of Sculptor. The model predictions are compared to 
   data from high- and medium-resolution spectra of giant stars in Sculptor 
   (see Sect.~\ref{sec:data} and figure captions for references). Data for a 
   few stars below [Fe/H]~= $-$3 dex seem to point to a non-negligible 
   dispersion which, in turn, would point to inhomogeneous chemical evolution 
   in the early galaxy. Since currently our model is not able to deal with 
   chemical inhomogeneities, we restrict our comparison to the metallicity 
   range $-$3~$<$ [Fe/H]~$< -$0.8. All abundance ratios are normalized to solar 
   values by Grevesse \& Sauval (1998), apart from Kirby et al. (2009) and 
   Shetrone et al. (2003), who use log($N_{\mathrm{Fe}}/N_{\mathrm{H}}$)~= 7.52 
   rather than 7.50 for the solar abundance of iron, and Geisler et al. (2005), 
   who adopt log($N_{\mathrm{O}}/N_{\mathrm{H}}$)~= 8.77 rather than 8.83 for the 
   solar abundance of oxygen. These differences of a few hundredths of dex are 
   smaller than quoted uncertainties of the measurements and were therefore 
   neglected.

   The ratio of $\alpha$ elements to iron is commonly believed to be a powerful 
   tracer of the timescale of formation of a stellar system, because of its 
   sensitivity to the ratio of short-lived SNII to long-lived SNIa progenitors. 
   During the earliest stages of the evolution, basically only SNeII contribute 
   to the chemical enrichment and, thus, high [$\alpha$/Fe] ratios are 
   observed. As soon as SNeIa start to dominate the iron production, a 
   \emph{`knee'} is produced in the [$\alpha$/Fe] versus [Fe/H] plot (Matteucci 
   2001, and references therein). Galaxies with low SFRs and/or that lose their 
   metals in a galactic wind will show (more or less clearly) the knee at 
   metallicities lower than galaxies with high SFRs that retain their metals. 
   Of particular interest as probes of the enrichment timescales are also those 
   elements whose yields are highly dependent on the metallicity of the parent 
   stars, such as manganese (Romano et al. 2011, and references therein). 


   \begin{figure}
   \psfig{figure=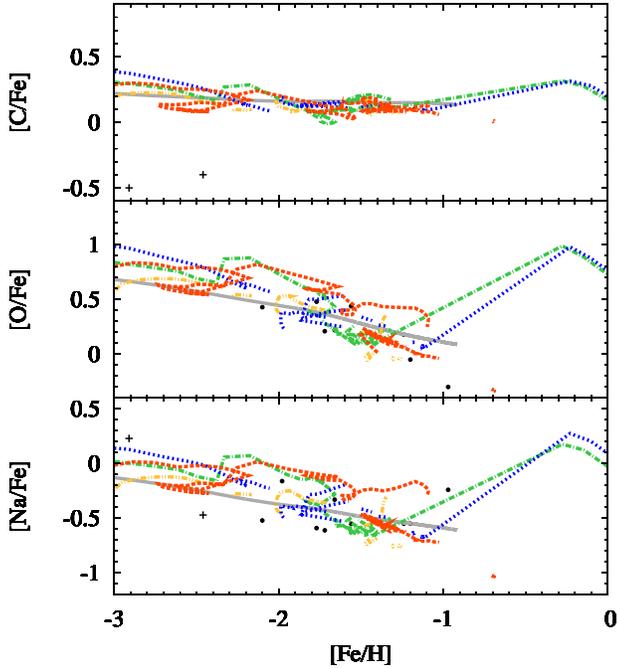,width=\columnwidth}
      \caption{ Same as Fig.~\ref{fig:mgsica}, for C, O and Na.
              }
         \label{fig:cona}
   \end{figure}


   Owing to its bursting star formation mode, with few short active phases 
   separated by long quiescent periods (see Fig.~\ref{fig:summary}, bottom left 
   box, yellow wall), model~Scl\,{\sevensize 2} displays the most `scrappy' 
   lines in Figs.~\ref{fig:mgsica} to \ref{fig:mnconi}, with many gaps in 
   [Fe/H]: the abundance of Fe, that is produced mainly by SNeIa on long 
   timescales, continues to grow in the ISM even if the star formation goes to 
   zero, unless the accretion of a substantial amount of metal-poor matter 
   dilutes the enriched medium. An important accretion of nearly unprocessed 
   gas actually happens several times in the evolution of 
   model~Scl\,{\sevensize 4} (notice the numerous right-to-left shifts in the 
   path followed by this model in Figs.~\ref{fig:mgsica} to \ref{fig:mnconi}). 
   Overall, our models are in reasonable agreement with the abundance data for 
   Sculptor, especially if considering that the yields of Sc and Ti from 
   massive stars are underestimated, while those of Cr are overestimated, in 
   the metallicity range probed by this study and that current SNIa models do 
   highly overestimate the production of Ni (see Romano et al. 2010a, and 
   references therein). In particular, model~Scl\,{\sevensize 4} is in 
   excellent agreement with the high-resolution data for Mg in Sculptor 
   (Fig.~\ref{fig:mgsica}, top panel) and nicely reproduces the striking 
   decreasing trend of [Mn/Fe] versus [Fe/H] found in this galaxy (North et al. 
   2012; see Fig.~\ref{fig:mnconi}, top panel).

   For the $\alpha$-element magnesium, model~Scl\,{\sevensize 1} (green 
   dot-dashed line in Fig.~\ref{fig:mgsica}, upper panel) predicts a knee 
   steeper than suggested by the high-resolution data in the [$\alpha$/Fe] 
   versus [Fe/H] plot, while model~Scl\,{\sevensize T} produces a curve that is 
   too flat (grey solid line in Fig.~\ref{fig:mgsica}, upper panel). Since all 
   model galaxies presented in this work share the same prescriptions about 
   SNIa progenitors and nucleosynthesis (see Sect.~\ref{sec:nuc}), we conclude 
   that the specific histories of star formation and mass assembly play a 
   crucial role in determining the exact shape of the [Mg/Fe]--[Fe/H] (and 
   [Mn/Fe]--[Fe/H]) relation in Sculptor.

   As already mentioned, at the lowest metallicities, $-$4~$<$ [Fe/H]~$< -$3, 
   the data show a significant dispersion. Stochastic sampling of the IMF as a 
   consequence of the low SFRs (Carigi \& Hernandez 2008; Cescutti 2008) may 
   be, at least partly, responsible for the observed scatter. Our model does 
   not incorporate yet inhomogeneous mixing of pockets of gas which are 
   enriched by a certain type of SN event. However, there is increasing 
   evidence that such inhomogeneous mixing exists in the early generations of 
   star formation in dwarf galaxies (e.g. Tafelmeyer et al. 2010; Venn et al. 
   2012). In the data set of very metal-poor Sculptor stars of Starkenburg et 
   al. (2013b), one star shows low $\alpha$- and heavy elements compared to 
   iron. This is consistent with a picture in which the star was born from a 
   SNIa enriched pocket (Marcolini et al. 2006). A stronger case of such a star 
   was discovered in the Carina dwarf galaxy (Venn et al. 2012). Efforts are 
   ongoing to include inhomogeneous mixing in our chemical evolution code.


   In Fig.~\ref{fig:cona} we show, from top to bottom, our predictions for 
   [C/Fe], [O/Fe] and [Na/Fe] versus [Fe/H] in Sculptor. These predictions need 
   to be confirmed (or disproved) by future observations. In fact, only sparse 
   data are available at present for these elements. As for carbon, we caution 
   that in giant stars dredge-up of CNO-processed material to the surface may 
   expose C-poor, N-rich matter and complicate the interpretation of the 
   abundances (Spite et al. 2005).

   \subsection{Major model uncertainties}
   \label{sec:unc}


   \begin{figure}
   \psfig{figure=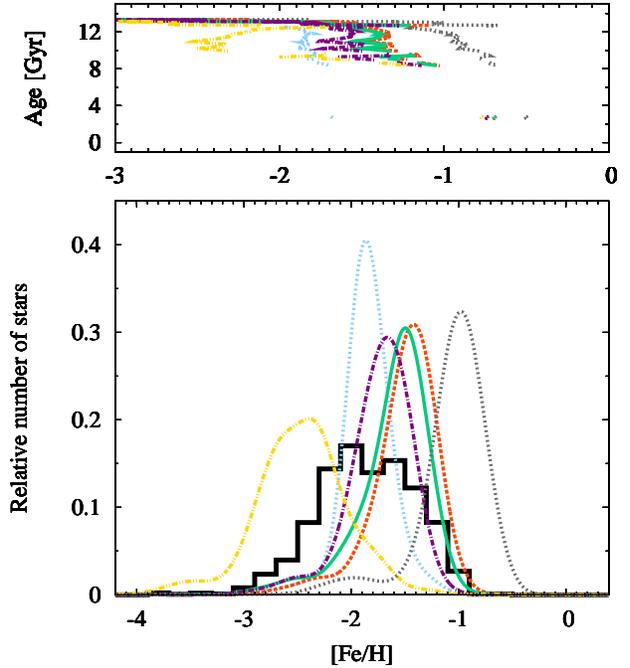,width=\columnwidth}
      \caption{ Age-metallicity relations (upper panel) and MDFs (bottom panel) 
        obtained with model Scl\,{\sevensize 4} with different prescriptions 
        about the IMF or the metal losses (yellow dot-dot-dashed, purple 
        dot-dashed, green solid, red dashed and grey dotted lines; see text) or 
        without including the contribution of SNeIa to the stellar 
        nucleosynthesis (sky-blue dotted lines). The theoretical MDFs have been 
        smoothed by a Gaussian function with a variance equal to the data 
        error, 0.15 dex. The black solid histogram is the observational MDF we 
        obtain by combining the DART CaT sample (Battaglia et al. 2008b; 
        Starkenburg et al. 2010) with the sample by Kirby et al. (2010), as 
        specified in Appendix~\ref{sec:A}.
              }
         \label{fig:mdf4}
   \end{figure}


   Up to now, we have compared the outputs of different models for Sculptor 
   ---a classical, `non-cosmological' one (model labelled Scl\,{\sevensize T}) 
   and four ones based on full cosmological simulations (models labelled 
   Scl\,{\sevensize 1}, Scl\,{\sevensize 2}, Scl\,{\sevensize 3} and 
   Scl\,{\sevensize 4}). The values of the parameters for 
   model~Scl\,{\sevensize T} are listed in Sect.~\ref{sec:classic}; they are 
   fixed mainly by the requirement of reproducing the SFH of Sculptor inferred 
   from the observations (de Boer et al. 2012), as well as its stellar MDF 
   (this work, Appendix~\ref{sec:A}). As for models~Scl\,{\sevensize 1}, 
   Scl\,{\sevensize 2}, Scl\,{\sevensize 3} and Scl\,{\sevensize 4}, the 
   results that we present in Sects.~\ref{sec:amrmdf} and \ref{sec:ratios} rest 
   on the following assumptions: (i) the hot ejecta of SNe of all types cool 
   and mix with the ISM on short time scales (i.e., shorter than the typical 
   time step for computation); (ii) the metals heated by SN explosions and 
   entrained in the outflow never re-enter star formation in the system; (iii) 
   the realization probability for SNIa events is the same as in our Galaxy. To 
   conclude our inspection of Sculptor-like model galaxies, we show in the 
   following how the predictions of model~Scl\,{\sevensize 4} change when 
   modifying these standard assumptions. Furthermore, we quantify the effects 
   of small variations in the IMF slope.

   \subsubsection{The role of SNeIa}
   \label{sec:snia}

\begin{figure}
   \psfig{figure=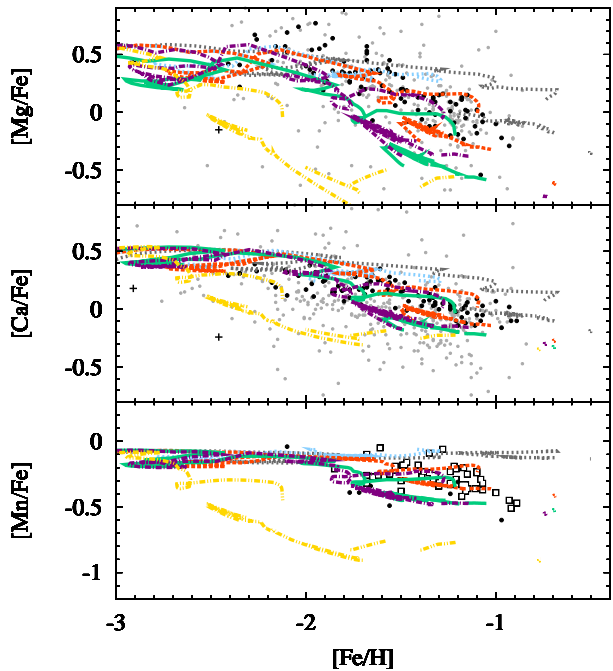,width=\columnwidth}
      \caption{ [X/Fe] versus [Fe/H] relations for Mg, Ca and Mn in Sculptor. 
        The predictions of model~Scl\,{\sevensize 4} with different 
        prescriptions about the IMF or the metal losses, or without including 
        SNeIa nucleosynthesis (lines, see text and caption to 
        Fig.~\ref{fig:mdf4}) are displayed and compared to the relevant data 
        (symbols, see Sect.~\ref{sec:data} and captions to 
        Figs.~\ref{fig:mgsica} and \ref{fig:mnconi} for references).
              }
         \label{fig:mgcamn4}
   \end{figure}

   In Figs.~\ref{fig:mdf4} and \ref{fig:mgcamn4}, we show the predictions 
   concerning the age-metallicity relation, MDF and [X/Fe] versus [Fe/H] 
   behaviour (for Mg, Ca and Mn; we select only elements with both reliable 
   nucleosynthesis prescriptions and homogeneous measurements from 
   high-resolution spectra in a large number of stars) of model 
   Scl\,{\sevensize 4} computed including the contribution from SNeIa (red 
   dashed lines; standard choice) and without SNeIa (sky-blue dotted lines). It 
   is immediately seen that without SNeIa the predicted MDF peaks towards lower 
   metallicities, in better agreeement with the observed one, but it narrows as 
   well, at variance with the observations. Furthermore, without SNeIa there is 
   no knee in the [$\alpha$/Fe] versus [Fe/H] plot and the observed decrease of 
   [Mn/Fe] for [Fe/H]~$> -$1.5 dex is not reproduced any more. Clearly, SNeIa 
   are of primary importance to reproduce the chemical features of stars in 
   Sculptor. This result is not new ---it has been shown several times in the 
   literature that SNeIa are required in order to accurately reproduce the 
   chemical compositions of galaxies. Nevertheless, we deem worth showing here 
   the results of model Scl\,{\sevensize 4} computed without SNeIa, to make it 
   clear that the lack of very metal-poor stars below [Fe/H]~$= -$2.3 dex, 
   plaguing all of our cosmologically-motivated models for Sculptor (see 
   Fig.~\ref{fig:mdf}, lower panel) is almost unrelated to the number of prompt 
   SNeIa exploding in the models (see Fig.~\ref{fig:mdf4}, lower panel; red 
   dashed versus sky-blue dotted lines). Our failure in reproducing the very 
   low-metallicity tail of the observed MDF would rather point to the need for 
   more diluting gas and/or a more efficient sink for metals during the 
   earliest phases of galaxy evolution.
   
   We stress here that, while in our post-processing code the contribution of 
   SNeIa to the nucleosynthesis is taken into account in detail (see 
   Sect.~\ref{sec:nuc}), in the adopted SAM SNeIa are not included, either in 
   the metal or energy budgets. Since their contribution to feedback processes 
   in galaxies may become important as well, especially during the late stages 
   of the evolution in case of protracted star formation (e.g. Bradamante et 
   al. 1998; Recchi et al. 2001), we caution that the results presented in this 
   paper could change when a detailed treatment of SNIa feedback is included in 
   the SAM.


   \subsubsection{The role of the IMF}

   Throughout this paper, we adopt an extrapolation of the Salpeter (1955) IMF, 
   defined in the mass range 0.1--100~M$_\odot$. This is a common choice in 
   chemical evolution studies. However, in order to investigate the stability 
   of our results against (plausible) IMF variations, we also run 
   model~Scl\,{\sevensize 4} by assuming a Chabrier-like IMF normalized to 
   unity in the 0.001--100~M$_\odot$ mass range and with $x =$~1.7 in the 
   exponential law for $m >$ 1~M$_\odot$ ($x =$~1.35 for Salpeter).

   We note that in the original paper, Chabrier (2003) quotes $x 
   =$~1.3\,$\pm$\,0.3 to take the observational errors into account. Here we 
   adopt an extreme ---but still empirically supported (see Kroupa 2001, 
   2012)--- value for the IMF slope. We do this in order to maximize the 
   differences in the model outputs with respect to our standard (Salpeter) 
   choice. With our particular choice of a Chabrier-like IMF, the predicted MDF 
   shows a negligible shift towards lower [Fe/H] values (green solid versus red 
   dashed lines; Fig.~\ref{fig:mdf4}, bottom panel). The predicted [X/Fe] 
   ratios are lowered (green solid versus red dashed lines; 
   Fig.~\ref{fig:mgcamn4}), but the effect is noticeable ($\sim$0.3 dex at 
   maximum) only for the elements originating mostly from SNeII, such as Mg.

   Overall, reasonable changes in the IMF produce secondary-order effects on 
   the predicted MDF of our model galaxies (see also Marconi et al. 1994). 
   However, the predicted abundance ratios may vary significantly, depending on 
   the elements.

   \subsubsection{The role of metal recycling through the hot phase}
   \label{sec:rec}

   Following Li et al. (2010) and Starkenburg et al. (2013a), 
   models~Scl\,{\sevensize 1}, Scl\,{\sevensize 2}, Scl\,{\sevensize 3} and 
   Scl\,{\sevensize 4} include a route to recycle the metals produced by dying 
   stars through the hot phase of a galaxy. In our standard model, a 
   substantial fraction [$F =$ 0.95 in equation~(\ref{eq:basic})] of metals is 
   ejected directly into the hot gas component and can be reincorporated in the 
   cold gas phase later on (the remainder goes directly in the cold phase). 
   Here we investigate further this assumption by studying two extreme cases: 
   (i) SN ejecta put in the hot phase are directly available for cooling again; 
   (ii) SN ejecta are subtracted from the cold gas phase forever. The yellow 
   (dot-dot-dashed) lines in Figs.~\ref{fig:mdf4} and \ref{fig:mgcamn4} show 
   the predictions of model~Scl\,{\sevensize 4} in case the metals deposited in 
   the hot phase are made to never re-enter regions of active star formation, 
   compared to the case in which they are immediately and fully reincorporated 
   (standard choice, red dashed lines). In case of inefficient metal recycling, 
   the peak of the theoretical MDF shifts towards lower metallicities and the 
   distribution broadens. One might be, thus, led to believe that the 
   best-fitting solution involves some fine tuning, with large fractions of the 
   newly-produced metals deposited directly in the hot gas phase never getting 
   back to regions of active star formation. However, a glance at 
   Fig.~\ref{fig:mgcamn4} (yellow dot-dot-dashed versus red dashed lines) 
   reveals that, in order to honour the constraints imposed by the trends of 
   the abundance ratios with metallicity, most of the stellar ejecta must 
   instead cool and mix with the neutral ISM forming the next generations of 
   stars!

   Apart from the injection of freshly produced metals from dying stars 
   directly in the hot component, another mode of metal removal from the 
   star-forming regions is active in our models, i.e. entrainement of (part of) 
   the ISM perturbed by SN explosions in the outflow [this is the ejected gas 
   component, see Fig.~\ref{fig:chart}; see also Sect.~\ref{sec:gas}, 
   equation~(\ref{eq:xi})]. In our standard scheme, the metals carried away by 
   the outflow are assumed to be definitively lost from the system. If, 
   instead, they are fully recycled through the hot gas phase, the theoretical 
   MDF is found to span the metallicity range $-$2.5~$<$ [Fe/H]~$< -$0.5 with a 
   peak at [Fe/H]~$-$1.0 dex, at variance with the observations; furthermore, 
   the knees in the theoretical [$\alpha$/Fe]--[Fe/H] relations are shifted 
   towards higher metallicities and flatter relations are predicted (grey 
   dotted lines in Figs.~\ref{fig:mdf4} and \ref{fig:mgcamn4}). This is because 
   the system can reach higher metallicities by the time SNeIa start to 
   contribute the bulk of their Fe to the ISM. Finally, the purple (dot-dashed) 
   curves in Figs.~\ref{fig:mdf4} and \ref{fig:mgcamn4} refer to 
   model~Scl\,{\sevensize 4} computed by assuming $\xi_{i}$~= 1 for H and He 
   and $\xi_{i}$~= 2 for metals in equation~(\ref{eq:xi}) (in the standard 
   case, red dashed lines, we set $\xi_{i}$~= 1 for all elements). This choice 
   is the analogous to the differential, metal-enriched wind hypothesis in 
   classical chemical evolution studies. With this choice, the peak of the 
   distribution shifts towards lower metallicities, but the model barely 
   matches the available abundance data. Setting $\xi_{i}$ to even higher 
   values for metals results in unacceptable theoretical [X/Fe] versus [Fe/H] 
   relations: in fact, the knee in the [$\alpha$/Fe] versus [Fe/H] plot moves 
   to [Fe/H]~$< -$2.3 dex, at variance with the observations, and a slope 
   steeper than observed is obtained for [Fe/H]~$> -$2 dex (models not shown in 
   Figs.~\ref{fig:mdf4} and \ref{fig:mgcamn4}, to avoid overcrowding).

   Summarizing: Accretion of pristine ---or nearly unpolluted--- gas, either 
   through infall or ingestion of small, gas-rich satellite systems, and 
   subsequent conversion of this gas into stars play a fundamental role in 
   determining the final chemical properties of galaxies. Having fixed the 
   histories of mass accretion and star formation by means of cosmological 
   simulations and a SAM of galaxy formation, the results presented in this 
   work are largely driven by the amount of metals that the galaxy is able to 
   lose ---or, better, to subtract from the cold, star-forming phase--- at any 
   time, \emph{as well as to the mode of the metal losses.} Indeed, it makes a 
   difference if the SN ejecta are vented out of the galaxy directly, i.e. 
   without interacting with the surroundings, or if some mixing with the 
   ambient medium is permitted before. We prefer a scenario in which a 
   significant dilution does occur: the simultaneous comparison of our model 
   results with both a well-defined observational MDF and high-quality 
   abundances in a large sample of stars allows us to significantly constrain 
   the parameter space of the model and to discriminate among different 
   possible evolutive scenarios.

   \section{Discussion}

   The idea of substantial mass loss on a galactic scale from starbursts in 
   small galaxies is not new. In particular, the existence of differential 
   winds, i.e. galactic winds in which heavier elements are vented out of the 
   galaxy more efficiently than lighter ones, was first hypothesized and 
   applied to the chemical evolution of generic dwarf galaxies by Pilyugin 
   (1993). Shortly after, Marconi et al. (1994) introduced the concept of 
   \emph{selective winds}, i.e. differential winds in which different metals 
   have different ejection efficiencies in dependence on the nature of the 
   parent stars (see also Recchi et al. 2001; Fujita et al. 2004; Romano et al. 
   2010b). Nowadays, metal-enriched outflows are often invoked by modellers to 
   reproduce the observed metallicity-luminosity relation of dwarf galaxies, as 
   well as detailed abundance data for specific objects, in the context of both 
   classical chemical evolution studies (e.g. Carigi et al. 2002; Lanfranchi \& 
   Matteucci 2003, 2004; Romano et al. 2006; Yin et al. 2011) and \emph{ab 
     initio} galaxy formation models (e.g. Salvadori et al. 2008; Calura \& 
   Menci 2009; Sawala et al. 2010). In the following, we discuss our findings 
   in comparison to recent theoretical studies dealing with the Sculptor dSph.

   In their numerical chemical evolution model for Sculptor, Lanfranchi \& 
   Matteucci (2003, 2004) have adopted the SFH inferred from the CMDs (Dolphin 
   2002) and imposed that the metals produced by SNe of all types are 
   efficiently removed by strong differential galactic winds. Notwithstanding 
   this, their Sculptor gets too rapidly relatively metal-rich and the 
   theoretical MDF (Lanfranchi \& Matteucci 2004, their figure~6) completely 
   lacks the most metal-poor stars observed in Sculptor. In our `cosmological' 
   models, we similarly miss the stars with [Fe/H]~$< -$2.5 dex (this work, 
   Fig.~\ref{fig:mdf}, lower panel). However, the shape of the [Mg/Fe] versus 
   [Fe/H] relation predicted by both Lanfranchi \& Matteucci (2004, their 
   figure~3) and ourselves (this work, Fig.~\ref{fig:mgsica}, upper panel) 
   agrees very well with observations of giant stars in Sculptor. We also note 
   that in Lanfranchi \& Matteucci's study the ratio of $\alpha$-elements to Fe 
   in the ISM of Sculptor is predicted to monotonically decrease in time (see 
   also the results of our classic model, grey lines in Fig.~\ref{fig:mgsica}). 
   This is essentially due to the monotonic behaviour of the assumed accretion 
   rate of pristine gas for star formation, leading to a one-to-one 
   age-metallicity relation for Sculptor. Our `cosmological' models, instead, 
   display more complicated AMRs, because of the much more complex histories of 
   mass assembly predicted by the underlying SAM. Thus they predict, within a 
   given galaxy, the existence of stars with the same metallicity, but with 
   different ages and, hence, with different abundance ratios. This can explain 
   a moderate degree of inhomogeneity (up to 0.3~dex) in the data. 
   Models~Scl\,{\sevensize 1} and Scl\,{\sevensize 3} also interestingly 
   predict that in a minority of metal-rich stars with [Fe/H]~$>$ 0 dex, the 
   abundance ratios are reset to the values reflecting type II SN 
   nucleosynthesis. This happens because of the strong star formation bursts 
   that end the evolution of these model galaxies some 8 Gyr ago (see 
   Fig.~\ref{fig:summary}): the gas is almost exhausted, the chemical imprints 
   of previous galactic evolution are washed out and we just see the 
   preponderant signature of the latest numerous core-collapse SNe. Although, 
   actually, no stars with [Fe/H]~$> -$0.8 dex are observed in Sculptor. In the 
   case of model Scl\,{\sevensize 3}, the last burst occurs after the galaxy 
   has become a satellite, in which case the physics of star formation has 
   become even more uncertain to predict ---it is, therefore, unclear if we can 
   exclude these model galaxies as good representatives of the Sculptor dwarf 
   spheroidal on the basis of a very small population originating in one 
   particular event. The mechanism discussed above is interesting as it could, 
   in principle, explain the existence of a Mg-rich population at intermediate 
   ages, as observed in the Carina dwarf galaxy (Lemasle et al. 2012). We plan 
   to deal with the Carina dSph, as well as other well-studied dwarf galaxies 
   of the Local Group, in a forthcoming paper of this series.
   
   More recently, Kirby et al. (2011a) have readdressed the issue of the 
   chemical evolution of Sculptor in the light of their new data and discussed 
   some simple analytic models for Sculptor, tailored to reproduce the 
   observational MDF they present elsewhere (Kirby et al. 2009, 2010). The 
   low-metallicity tail of the observed distribution is reproduced by their 
   simple models, but the peak of the theoretical distributions is located at 
   [Fe/H]~=$-$1.6. This is in reasonable agreement with the MDF derived from 
   observations of stars in the inner $\sim$0.2$\degr$ region of Sculptor 
   (Kirby et al. 2009, 2010), but in poorer agreement with the MDF 
   representative of the full Sculptor galaxy derived in this paper (see 
   Appendix~\ref{sec:A}). Overall, their theoretical MDFs are quite similar to 
   the one we obtain in the framework of the classical chemical evolution model 
   discussed in this work (model labelled Scl\,{\sevensize T} in previous 
   sections).

   In general, it seems difficult to reproduce the correct fraction of stars 
   with $-$2.5~$<$ [Fe/H] $< -$1.5 observed in Sculptor, and this problem 
   pertains to both classical and `cosmological' models. Possible solutions 
   include: (i) the implementation of Population~III stars (with highly 
   uncertain yields and IMF) in the models; (ii) a favoured black hole 
   formation as the outcome of massive star evolution at low metallicities 
   (with consequent sink of metals); (iii) a lower rate of occurrence of SNIa 
   events at low metallicities (and/or in dense systems) and (iv) assembly from 
   small subunits where the star formation is strongly suppressed in the early 
   stages of galaxy evolution. In a forthcoming paper (Romano et al., in 
   preparation), we will deal with points (i) to (iii). Points (i), (ii) and 
   (iv) also provide possible solutions to the problem of the low fractions of 
   stars below [Fe/H]~= $-$2.3 dex predicted by models Scl\,{\sevensize 1}, 
   Scl\,{\sevensize 2}, Scl\,{\sevensize 3} and Scl\,{\sevensize 4}. Revaz \& 
   Jablonka (2012) obtain a better fit to the low-metallicity wing of the 
   observed Sculptor MDF (as well as a good fit to the observed [Mg/Fe] versus 
   [Fe/H] relation) through a smoothed particle hydrodynamics code. Yet, their 
   Sculptor model needs to have its star formation artificially stopped. 
   Moreover, it retains a large amount of gas, $\mathscr{M}_{\mathrm{cold\,gas}}$ 
   = 1.9~$\times$ 10$^7$ M$_\odot$ at the present time, to be compared with 
   values one to two orders of magnitude lower for most of our models (see 
   Table~\ref{tab:mod}, third column). On the observational side, Grcevich \& 
   Putman (2009) have shown that $\mathscr{M}_{\mathrm{HI}}$~= 2.34~$\times$ 
   10$^5$ M$_\odot$ of neutral hydrogen are possibly associated to the Sculptor 
   dSph, but the detection is ambiguous.

   Based on their modelling, Kirby et al. (2011b) have estimated that Local 
   Group dSphs have lost from 96 to more than 99 per cent of the metals their 
   stars manufactured. They suggest that gas outflows carried away most of the 
   metals produced by these dSphs. For the post-starburst galaxy NGC\,1569, 
   there is indeed direct evidence (from X-ray spectral fit to several $\alpha$ 
   elements; Martin et al. 2002) that the galaxy is losing nearly all of the 
   metals it has produced in the latest starburst. However, three-dimensional 
   hydrodynamic simulations by Marcolini et al. (2006) exclude that local dSphs 
   got rid of their gas by internal mechanisms such as galactic winds. 
   According to those authors, the evolution towards gas-poor systems would 
   result from external mechanisms (ram pressure stripping and/or tidal 
   interactions with the Milky Way). In their models, the SN ejecta remain 
   gravitationally bound to the parent system. Yet, only a small fraction (less 
   than 18 per cent) of it lies in the region where the star formation is 
   active, which avoids the production of metal-rich stars. In Marcolini et 
   al.'s (2006) simulations, the knee in the [$\alpha$/Fe] versus [Fe/H] 
   diagram is due to the inhomogeneous distribution of the SNIa ejecta, rather 
   than to the combined effect of the time delays with which SNeIa restore the 
   bulk of their iron to the ISM and the onset of a galactic wind, as in 
   classical chemical evolution studies (and this work). However, a significant 
   scatter in the [O/Fe] ratios is predicted at relatively high metallicities, 
   that is not confirmed observationally. In fact, high-resolution 
   spectroscopic data ---and medium-resolution data, taking into account the 
   larger uncertainties in these data--- for hundreds of stars in Sculptor 
   point rather to a remarkable homogeneity of the ISM during most of the 
   evolution of the system (Kirby et al. 2009, 2010; Tolstoy et al. 2009; North 
   et al. 2012; Hill et al., in preparation). 

   In the works by Li et al. (2010) and Starkenburg et al. (2013a), the 
   following feedback scheme is adopted: (i) 95 per cent of SN ejecta goes 
   directly into the hot gas phase (the remainder 5 per cent instantly pollutes 
   the neutral ISM); (ii) metals in the ISM perturbed by SN explosions are put 
   in an ejected component; they can re-enter the cold gas phase if recycled 
   through the hot gas. In our models, we assume the cooling flow (grossly, H 
   plus He) dictated by the cosmological simulations and SAM. The metallicity 
   of the flow is further investigated: (i) SN ejecta in the hot gas are either 
   put all straight into the cold gas phase (standard choice) or made never 
   enter the cold gas phase; (ii) metals in the ejected component are either 
   definitively lost from the system (standard choice) or fully recycled 
   through the hot gas phase. Our standard scheme, namely the one in which SN 
   ejecta mix instantaneously with the surrounding ISM, but the metals in the 
   ejected component are lost, does not have a physical motivation. But we do 
   find that it reproduces the observed chemical properties of Sculptor' stars. 
   Full hydrodynamical simulations are needed in order to deal properly with 
   issues such as the interaction of the metal-rich matter processed by the 
   starburst with the ambient medium and the metal losses from the galaxy (see 
   Recchi \& Hensler 2013, for a recent review of those studies). Our feedback 
   scheme is independent of time and geometry of the system; but it is likely 
   that the efficiency of metal removal from the star forming regions varies 
   with time, in dependence of the mass and size of the star-forming regions 
   (see, e.g., Tenorio-Tagle et al. 2007; W{\"u}nsch et al. 2011). The 
   development of winds and the fate of metals also depend on the geometry of 
   the system: models with the same baryonic mass and SFH retain more or less 
   metals, depending on their degree of flattening (see Recchi \& Hensler 
   2013). Our code does not tackle the complex physics behind the 
   circulation/loss of metals within/from the galaxy. On the other hand, thanks 
   to its simplifications, it runs extremely fast, which makes it feasible to 
   compute many hundreds of models and fully explore the parameter space.

   Similarly to us, Calura \& Menci (2009) also performed a post-processing of 
   cosmological simulations for detailed chemistry and proposed that the 
   realization probability of SNeIa in Local Group dwarf galaxies must be lower 
   than in spirals such as the Milky Way. This assumption, joint to the 
   adoption of strongly metal-enhanced outflows, allows them to reproduce the 
   mass-metallicity relation of local dwarfs. However, they run models for 
   generic dwarfs, while we are focusing here on a particular object, the 
   Sculptor dSph. When we run models with lower values of the ${\mathscr B}$ 
   parameter (see Sect.~\ref{sec:gce}) for Sculptor, we end up with theoretical 
   MDFs narrower than observed. Furthermore, with this assumption we can not 
   reproduce the steep knee in the [Mg/Fe] versus [Fe/H] relation that 
   characterizes Sculptor's stars. However, though we do not favour a low value 
   for ${\mathscr B}$ during the full evolution of Sculptor, we can not exclude 
   a lower probability for SNIa events during the relatively early phases of 
   Sculptor formation (see discussion in Sect.~\ref{sec:snia} and this section, 
   forth paragraph).

   We have already noticed a potential problem related to post-processing, 
   namely, the differences in the cold gas masses computed with or without IRA 
   (see Sect.~\ref{sec:gas}). However, we have checked these are small (within 
   a few per cent) in our models. Another issue is that of the lack of feedback 
   from SNeIa in the adopted SAM. SNeIa keep exploding even if the star 
   formation goes to zero. Therefore, while their effect is negligible when the 
   star formation is active (because of the prevailing SNeII), one might expect 
   they can play a role in keeping hot the ISM during halts in star formation 
   activity. The final effect could be that of suppressing some of the late 
   episodes of star formation in the models. Another potential problem of using 
   a post-processing procedure is that the cooling curves used are metallicity 
   dependent (Sutherland \& Dopita 1993) and as the metallicities obtained in 
   the post-processing are not necessarily identical to the SAM, this might 
   lead to inconsistencies. However, we note that in the cases described here 
   the metallicity distribution function obtained by the SAM and the 
   post-processing chemical code are quite similar, therefore we do not expect 
   the cooling processes to change very significantly. In the context of this 
   work, we can only discuss such issues in a qualitative way. For a more 
   robust, quantitative discussion, one should implement the chemical evolution 
   equations in the SAM and compare the outputs of the self-consistent model 
   with the results obtained with the post-processing technique. This is 
   however beyond the scope of this paper. Moreover, we like to note that 
   implementing the full set of the chemical evolution equations in the SAM has 
   its own problems: the computational time strongly increases because we must 
   follow the evolution of each chemical species in each of the progenitors 
   along the merger tree.

   \section{Summary and conclusions}

   In this paper, we present a new chemical evolution model suited to follow 
   the chemical evolution of dwarf galaxies in a full cosmological approach. 
   The code adopts the histories of mass assembly and star formation derived 
   from first principles by means of a hierarchical SAM and accounts for the 
   contribution to the chemical enrichment from several stellar sources, 
   namely, low- and intermediate-mass stars, massive stars and SNeIa. The 
   lifetimes of stars of different initial masses are taken into account in 
   detail, as is the distribution of the delay times for SNIa explosions. We 
   adopt metallicity-dependent stellar yields that satisfactorily reproduce the 
   trends of several abundance ratios with metallicity in the Milky Way. All of 
   this is mandatory to study the abundance ratios of elements that have 
   different stellar progenitors, which are precious diagnostics of the 
   time-scales of structure formation and evolution. 

   In this study, we focus on the Sculptor dSph, for which high-quality data 
   exist for a large number of stars, as a test bed for our model. We adopted 
   mass assembly, star formation histories and gas flows for four different 
   Sculptor-like models from the catalogue of satellite galaxies generated by 
   the implementation of a version of the Munich SAM on the high-resolution 
   Aquarius cosmological simulations (Starkenburg et al. 2013a) as a backbone 
   for the post-processing chemical evolution code.

   We find that, once a specific path is chosen for the formation of a dwarf 
   galaxy inside a merging hierarchy of dark matter haloes, the shape of the 
   MDF, as well as the behaviour of the abundance ratios as functions of [Fe/H] 
   in the system, are dictated primarily by the occurrence and strength of 
   those physical processes able to remove\footnote{Gas removal can result from 
     internal (SN feedback, leading to a blow-out phase) or external (ram 
     pressure stripping and/or tidal interactions by the Milky Way) mechanisms, 
     or both, in our models.} a large fraction of the metals synthesized by the 
   stars from the regions where the star formation occurs. From our results we 
   conclude that in particular the tracks of [Mg/Fe] and [Mn/Fe] versus [Fe/H] 
   are powerful tracers of the mass assembly and star formation history of a 
   galaxy.

   All of our model galaxies have some problems in reproducing the whole body 
   of abundance data available for the Sculptor dSph. This is true for our 
   classical model, as well as for the `cosmological' ones. For instance, none 
   of the models matches the observed fraction of stars with $-$2.5~$<$ 
   [Fe/H]~$< -$2. The cosmologically-motivated models also miss most of the 
   stars that are found below [Fe/H]~$< -$2.5. The problem we face is an 
   `inverse' G-dwarf problem, in the sense that we are severely underestimating 
   the number of low-metallicity stars in the system. Possible solutions could 
   involve higher efficiencies of metal losses during the early phases of 
   galaxy formation, a diminished metal production from very metal-poor stars, 
   or the presence of more gas to dilute the metals at early stages, without a 
   corresponding increase in the star formation rate. In the metallicity range 
   $-$3~$<$ [Fe/H]~$< -$0.8, all the models fit the available abundance data 
   reasonably well.

   As the infall of gas is fixed in the cosmologically motivated models, we can 
   use our results to constrain the loss of metals needed and the mode of these 
   metal losses. We find that the models prefer a significant dilution, in 
   which supernova gas first interacts with the surrounding gas before it is 
   lost forever from the star forming medium of the galaxy.

   \section*{Acknowledgements}

   The authors are indebted to Francesca Matteucci for having provided an 
   earlier version of the classical chemical evolution code and to Vanessa Hill 
   for having provided her data in advance of publication. The authors are 
   indebted to the Aquarius Simulations Consortium; in particular, they are 
   grateful to Gabriella De Lucia, Amina Helmi and Yang-Shyang Li for their 
   role in developing the semi-analytic model of galaxy formation used in this 
   paper. Several colleagues read an earlier version of this paper and provided 
   insightful comments; they are: Francesco Calura, Thomas de Boer, Gabriella 
   De Lucia, Amina Helmi, Vanessa Hill, Francesca Matteucci, Alan McConnachie, 
   Monica Tosi and Kim Venn. The paper also greatly benefited from comments by 
   an anonymous referee. DR and ES thank the International Space Science 
   Institute (ISSI, Bern, CH) for support of the teams ``Defining the full 
   life-cycle of dwarf galaxy evolution: the Local Universe as a template'' and 
   ``The Evolution of the First Stars in Dwarf Galaxies''. DR acknowledges 
   partial financial support from PRIN INAF~2009, project ``Formation and Early 
   Evolution of Massive Star Clusters'', and PRIN MIUR~2010--2011, project 
   ``The Chemical and Dynamical Evolution of the Milky Way and Local Group 
   Galaxies'', prot.~2010LY5N2T. ES is supported by the Canadian Institute for 
   Advanced Research (CIfAR) Junior Academy and by a Canadian Institute for 
   Theoretical Astrophysics (CITA) National Fellowship.

   \appendix

   \section{The Sculptor MDF}
   \label{sec:A}

   Both the DART project (Tolstoy et al. 2006; Helmi et al. 2006; Battaglia et 
   al. 2008b; Starkenburg et al. 2010) and Kirby et al. (2009, 2010) provide 
   spectroscopic datasets of large samples of individual RGB stars within the 
   Sculptor dwarf galaxy. However, since each dataset has its own observational 
   biases with respect to its radial extent and depth, the resulting MDFs are 
   quite different. In this Appendix, we strive to combine the two datasets 
   available in order to remove the observational biases as much as possible 
   and obtain a MDF that is representing all stellar populations of the 
   Sculptor dwarf galaxy. 

   Firstly, the sample from the DART project provides metallicities derived 
   through measurements of the Ca\,{\sevensize II} triplet line strengths for 
   over 600 stars in Sculptor out to $\sim$1.5$\degr$ in elliptical radii 
   (Tolstoy et al. 2006; Helmi et al. 2006; Battaglia et al. 2008b). The MDF 
   for the sample is derived using the latest calibration of Ca\,{\sevensize 
     II} triplet line strengths to [Fe/H] (Starkenburg et al. 2010). 
   Additionally, a high-resolution multi-object study was carried out in the 
   centre of the galaxy by DART (Hill et al., in preparation). The [Fe/H] 
   values derived from individual Fe lines in this high-resolution study do 
   agree well with the measurements of [Fe/H] via the strong Ca\,{\sevensize 
     II} triplet lines for the stars overlapping in both the low- and 
   high-resolution samples (Battaglia et al. 2008b; Starkenburg et al. 2010). 
   Although the DART dataset goes down to faint magnitudes ($V \sim$20) in the 
   outskirts of the galaxy, the central sample only fully covers the brightest 
   $\sim$1 magnitude below the tip of the RGB (until $V \sim$18). As shown by 
   de Boer et al. (2012), this results in a bias in the MDF in this region of 
   the galaxy, as low-metallicity stellar populations are underrepresented on 
   the upper RGB. They conclude that a coverage down to $V$~=~19.5 is at least 
   required to obtain an unbiased MDF.

   Secondly, Kirby et al. (2009, 2010) present a MDF from spectroscopic 
   observations of nearly 400 RGB stars distributed within $\sim$0.2$\degr$ in 
   elliptical radii. The resolution of their study is comparable to the 
   low-resolution DART study, but their spectra cover a larger wavelength 
   region. To derive [Fe/H] and abundance ratios for several $\alpha$ elements 
   they apply a spectral synthesis technique (see Kirby et al. 2008 for 
   details). They find very good agreement when comparing results for stars 
   overlapping with the high-resolution sample of Hill et al. (in preparation), 
   presented in Battaglia et al. (2008b). Their MDF is also shown to be 
   comparable to the DART MDF if the same radial and magnitude cuts are applied 
   to both samples (see de Boer et al. 2012, their figure~6, where they show 
   that for $r_{\mathrm{ell}} \le$ 0.2$\degr$ and truncation at $V \simeq$18 the 
   two MDFs are the same within the respective errorbars). The stars studied in 
   Kirby et al. (2009, 2010) go down to sufficiently faint magnitudes to expect 
   an unbiased sample in depth. However, an observational bias is expected 
   because of the limited radial extent of their survey. The Sculptor dwarf 
   galaxy is known to have a strong metallicity gradient with radius (Tolstoy 
   et al. 2004; Westfall et al. 2006), as well as an age gradient (de Boer et 
   al. 2012). The sample from Kirby et al. (2009) covers only the inner parts 
   of the galaxy and, therefore, undersamples the older and more metal-poor 
   stellar population. 


   \begin{figure}
   \psfig{figure=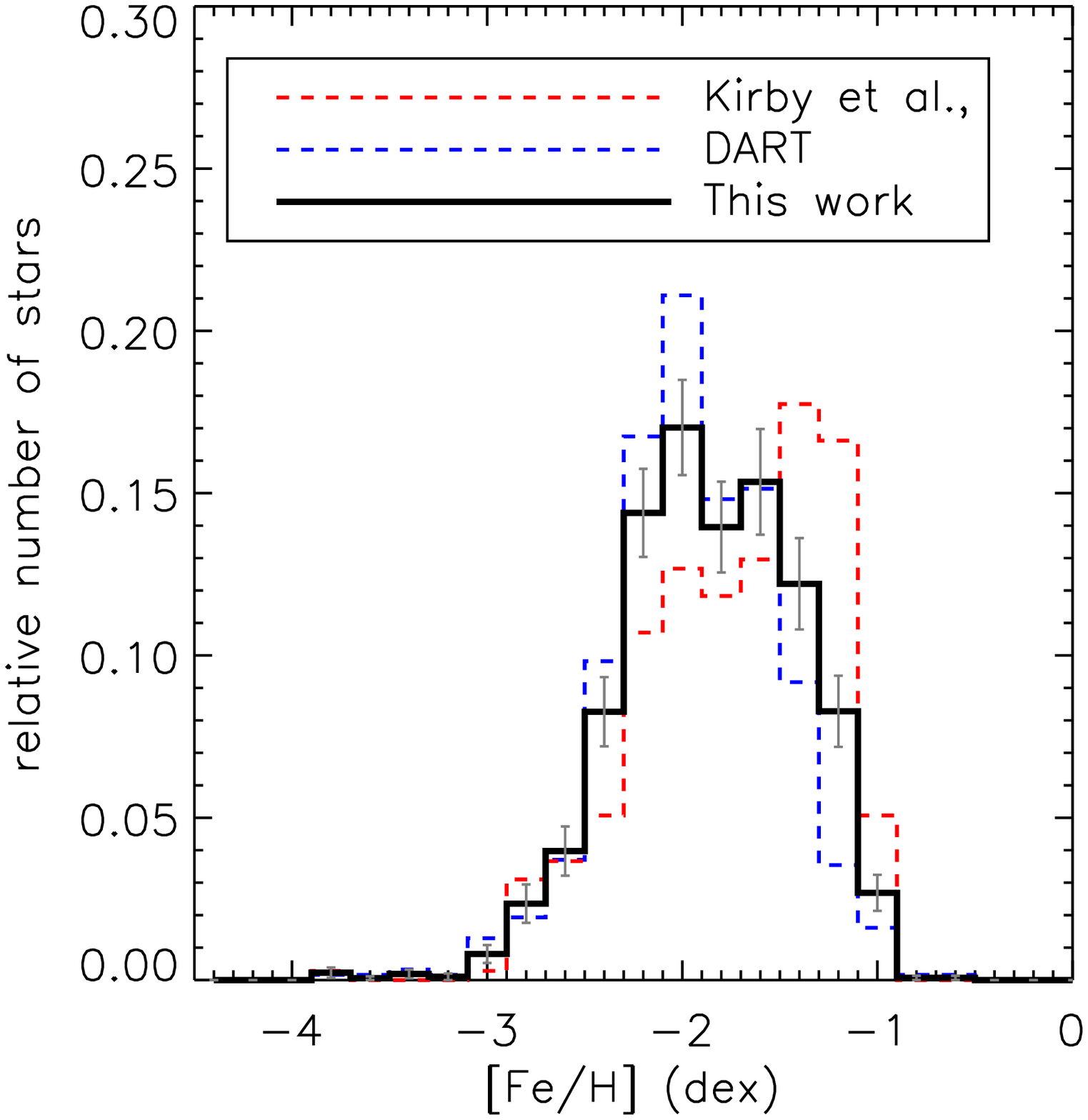,width=\columnwidth}
      \caption{The MDFs from Kirby et al. (2009; red dashed line) and the DART 
        collaboration (see Tolstoy et al. 2006; Helmi et al. 2006; Battaglia et 
        al. 2008b; Starkenburg et al. 2010; blue dashed line), from which the 
        MDF as used in this work has been derived (see text for details). The 
        final MDF is shown as the thick black line. Poissonian errorbars are 
        also shown.}
      \label{fig:appx}
    \end{figure}



   \begin{table}
     \caption{ Tabulated MDF for the final sample of about 1000 stars presented 
       in this work. This is shown as the thick black line in 
       Fig.~\ref{fig:appx}.}
     \begin{tabular}{rcc@{\hspace{0.9cm}}rcc}
       \hline
       [Fe/H]   & Rel. \#  & Error    & [Fe/H] & Rel. \#  & Error    \\
       (dex)    &          &          & (dex)  &          &          \\
       (1)      & (2)      & (3)      & (4)    & (5)      & (6)      \\
       \hline
       $-$4.4   & 0.00     & 0.00     & $-$1.8 & 1.39e-01 & 1.40e-02 \\
       $-$4.2   & 0.00     & 0.00     & $-$1.6 & 1.53e-01 & 1.63e-02 \\
       $-$4.0   & 0.00     & 0.00     & $-$1.4 & 1.22e-01 & 1.41e-02 \\
       $-$3.8   & 2.27e-03 & 1.61e-03 & $-$1.2 & 8.28e-02 & 1.10e-02 \\
       $-$3.6   & 5.77e-04 & 5.77e-04 & $-$1.0 & 2.69e-02 & 5.55e-03 \\
       $-$3.4   & 1.88e-03 & 1.34e-03 & $-$0.8 & 6.73e-04 & 6.73e-04 \\
       $-$3.2   & 1.01e-03 & 1.01e-03 & $-$0.6 & 6.73e-04 & 6.73e-04 \\
       $-$3.0   & 8.02e-03 & 2.76e-03 & $-$0.4 & 0.00     & 0.00     \\
       $-$2.8   & 2.35e-02 & 5.91e-03 & $-$0.2 & 0.00     & 0.00     \\
       $-$2.6   & 3.97e-02 & 7.57e-03 &    0.0 & 0.00     & 0.00     \\
       $-$2.4   & 8.26e-02 & 1.07e-02 &    0.2 & 0.00     & 0.00     \\
       $-$2.2   & 1.44e-01 & 1.36e-02 &    0.4 & 0.00     & 0.00     \\
       $-$2.0   & 1.70e-01 & 1.47e-02 &        &          &          \\
       \hline
       \end{tabular}
       \label{tab:appx}

       \medskip
       \emph{Note.} Different columns list: (1) and (4) the central value of 
       the metallicity bin; (2) and (5) the relative number of stars within the 
       metallicity bin; (3) and (6) the Poissonian error in the relative number 
       of stars.\\
   \end{table}


   In Fig.~\ref{fig:appx} the MDFs from DART and Kirby et al. (2009) are both 
   shown, as blue and red dashed lines, respectively. Both MDFs shown here are 
   truncated at $V$~=~20 to ensure that both datasets used are equally deep. 
   Additionally, this magnitude cut-off gets rid of the worst S/N data. 
   Clearly, the shapes and peaks of the two MDFs do not agree. The apparent 
   bimodality in the observed MDF from Kirby et al. (2009) present as two 
   peaks at [Fe/H]~= $-$2.1 and $-$1.3 dex (of which the more metal-rich is the 
   dominant) is not seen in the larger and more radially extended DART sample, 
   which peaks at lower metallicity. To create a more homogeneous sample, we 
   recalculate the elliptical radius for each star in either sample using the 
   best-fit parameters from the photometric study of de Boer et al. (2011). 
   Subsequently we bin both samples in bins of 0.05$\degr$ in elliptical radius 
   and construct a separate MDF for each bin. We decided to use solely the 
   sample of Kirby et al. (2009) as a representant sample for the inner regions 
   of the galaxy ($r_{\mathrm{ell}} <$0.2$\degr$) due to its greater 
   observational depth and thus completeness, while we use the DART sample to 
   represent the outer regions (0.2$\degr$$< r_{\mathrm{ell}} <$1.5$\degr$). 
   Following Battaglia et al. (2008a) we represent the surface density of RGB 
   stars in the Sculptor dSph by a combined Plummer and Sersic profile with 
   half-light radii of 15.1 and 8.6 arcmins, respectively. By multiplying the 
   surface density and the area of each bin in elliptical radius, we obtain the 
   weight each bin contributes to the galaxy. Normalizing these weights we find 
   that the relative contribution from the inner bins ($r_{\mathrm{ell}} 
   <$0.2$\degr$) is roughly half of the total (48 per cent). The final relative 
   MDF, shown in Fig.~\ref{fig:appx} (entries are listed in 
   Table~\ref{tab:appx}), consists of all MDFs of each radius bin multiplied 
   with their relative weight. Throughout this work we will use this final MDF 
   as a representative MDF for the full Sculptor galaxy and as such compare it 
   directly to our models. 

   \bsp

   \label{lastpage}


\begin{thebibliography}{90}
\bibitem{}
Arrigoni M., Trager S. C., Somerville R. S., Gibson B. K., 2010, MNRAS, 402, 173
\bibitem{}
Battaglia G., Helmi A., Tolstoy E., Irwin M., Hill V., Jablonka P., 2008a, ApJ, 
681, L13
\bibitem{}
Battaglia G., Irwin M., Tolstoy E., Hill V., Helmi A., Letarte B., Jablonka P., 
2008b, MNRAS, 383, 183
\bibitem{}
Battaglia G., Helmi A., Morrison H., Harding P., Olszewski E. W., Mateo M., 
Freeman K. C., Norris J., Shectman S. A., 2005, MNRAS, 364, 433
\bibitem{}
Bennett C. L., et al., 2012, preprint (arXiv:1212.5225)
\bibitem{}
Bradamante F., Matteucci F., D'Ercole A., 1998, A\&A, 337, 338
\bibitem{}
Brook C. B., et al., 2012, MNRAS, 426, 690
\bibitem{}
Calura F., Menci N., 2009, MNRAS, 400, 1347
\bibitem{}
Cannon J. M., McClure-Griffiths N. M., Skillman E. D., C{\^o}t{\'e} S., 2004, 
ApJ, 608, 768
\bibitem{}
Carigi L., Hernandez X., 2008, MNRAS, 390, 582
\bibitem{}
Carigi L., Col{\'i}n P., Peimbert M., Sarmiento A., 1995, ApJ, 445, 98
\bibitem{}
Carigi L., Hernandez X., Gilmore G., 2002, MNRAS, 334, 117
\bibitem{}
Cescutti G., 2008, A\&A, 481, 691
\bibitem{}
Cescutti G., Matteucci F., Lanfranchi G. A., McWilliam A., 2008, A\&A, 491, 401
\bibitem{}
Chabrier G., 2003, PASP, 115, 763
\bibitem{}
Chiosi C., Matteucci F., 1982, A\&A, 110, 54
\bibitem{}
Croton D. J., et al., 2006, MNRAS, 365, 11
\bibitem{}
Da Costa G. S., 1984, ApJ, 285, 483
\bibitem{}
de Boer T. J. L., 2012, PhD thesis, Kapteyn Astronomical Institute, Univ.
Groningen
\bibitem{}
de Boer T. J. L., et al., 2011, A\&A, 528, A119
\bibitem{}
de Boer T. J. L., Tolstoy E., Hill V., Saha A., Olsen K., Starkenburg E., 
Lemasle B., Irwin M. J., Battaglia G., 2012, A\&A, 539, A103
\bibitem{}
Dekel A., Silk J., 1986, ApJ, 303, 39
\bibitem{}
De Lucia G., Blaizot J., 2007, MNRAS, 375, 2
\bibitem{}
De Lucia G., Helmi A., 2008, MNRAS, 391, 14
\bibitem{}
De Lucia G., Kauffmann G., White S. D. M., 2004, MNRAS, 349, 1101
\bibitem{}
D'Ercole A., Brighenti F., 1999, MNRAS, 309, 941
\bibitem{}
Dolphin A. E., 2002, MNRAS, 332, 91
\bibitem{}
Ferrara A., Tolstoy E., 2000, MNRAS, 313, 291
\bibitem{}
Font A. S., McCarthy I. G., Crain R. A., Theuns T., Schaye J., Wiersma R. P. 
C., Dalla Vecchia C., 2011, MNRAS, 416, 2802
\bibitem{}
Frebel A., Kirby E. N., Simon J. D., 2010, Nature, 464, 72
\bibitem{}
Fujita A., Mac Low M.-M., Ferrara A., Meiksin A., 2004, ApJ, 613, 159
\bibitem{}
Geisler D., Smith V. V., Wallerstein G., Gonzalez G., Charbonnel C., 2005, AJ, 
129, 1428
\bibitem{}
Goswami A., Prantzos N., 2000, A\&A, 359, 191
\bibitem{}
Grcevich J., Putman M. E., 2009, ApJ, 696, 385
\bibitem{}
Greggio L., 2005, A\&A, 441, 1055
\bibitem{}
Greggio L., Renzini A., 1983, A\&A, 118, 217
\bibitem{}
Grevesse N., Sauval A. J., 1998, Space Sci. Rev., 85, 161
\bibitem{}
Grieco V., Matteucci F., Pipino A., Cescutti G., 2012, A\&A, 548, A60
\bibitem{}
Guo Q., White S., Li C., Boylan-Kolchin M., 2010, MNRAS, 404, 1111
\bibitem{}
Guo Q., White S., Angulo R. E., Henriques B., Lemson G., Boylan-Kolchin M., 
Thomas P., Short C., 2013, MNRAS, 428, 1351
\bibitem{}
Heckman T. M., Sembach K. R., Meurer G. R., Strickland D. K., Martin C. L., 
Calzetti D., Leitherer C., 2001, ApJ, 554, 1021
\bibitem{}
Helmi A., et al., 2006, ApJ, 651, L121
\bibitem{}
Hunter D. A., Elmegreen B. G., 2004, AJ, 128, 2170
\bibitem{}
Irwin M., Hatzidimitriou D., 1995, MNRAS, 277, 1354
\bibitem{}
Iwamoto K., Brachwitz F., Nomoto K., Kishimoto N., Umeda H., Hix W. R., 
Thielemann F.-K., 1999, ApJS, 125, 439
\bibitem{}
James B. L., Tsamis Y. G., Barlow M. J., 2010, MNRAS, 401, 759
\bibitem{}
Kauffmann G., Colberg J. M., Diaferio A., White S. D. M., 1999, MNRAS, 307, 529
\bibitem{}
Kennicutt R. C., Jr., 1989, ApJ, 344, 685
\bibitem{}
Kirby E. N., Guhathakurta P., Sneden C., 2008, ApJ, 682, 1217
\bibitem{}
Kirby E. N., Martin C. L., Finlator K., 2011b, ApJ, 742, L25
\bibitem{}
Kirby E. N., Guhathakurta P., Bolte M., Sneden C., Geha M. C., 2009, ApJ, 705, 
328
\bibitem{}
Kirby E. N., Lanfranchi G. A., Simon J. D., Cohen J. G., Guhathakurta P., 
2011a, ApJ, 727, 78
\bibitem{}
Kirby E. N., et al., 2010, ApJS, 191, 352
\bibitem{}
Kobayashi C., Nomoto K., 2009, ApJ, 707, 1466
\bibitem{}
Kobayashi C., Tsujimoto T., Nomoto K., Hachisu I., Kato M., 1998, ApJ, 503, L155
\bibitem{}
Kobulnicky H. A., Skillman E. D., 2008, AJ, 135, 527
\bibitem{}
Komatsu E., et al., 2011 ApJS, 192, 18
\bibitem{}
Kroupa P., 2001, MNRAS, 322, 231
\bibitem{}
Kroupa P., 2012, in Stamatellos D., Goodwin S., Ward-Thompson D., eds, The 
Labyrinth of Star Formation, Springer, in press (arXiv:1210.1211)
\bibitem{}
Lanfranchi G. A., Matteucci F., 2003, MNRAS, 345, 71
\bibitem{}
Lanfranchi G. A., Matteucci F., 2004, MNRAS, 351, 1338
\bibitem{}
Larson R. B., 1974, MNRAS, 169, 229
\bibitem{}
Larson R. B., 1976, MNRAS, 176, 31
\bibitem{}
Lemasle B., Hill V., Tolstoy E., Venn K. A., Shetrone M. D., Irwin M. J., de 
Boer T. J. L., Starkenburg E., Salvadori S., 2012, A\&A, 538, A100
\bibitem{}
Li W., Chornock R., Leaman J., Filippenko A. V., Poznanski D., Wang X., 
Ganeshalingam M., Mannucci F., 2011, MNRAS, 412, 1473
\bibitem{}
Li Y.-S., White S. D. M., 2008, MNRAS, 384, 1459
\bibitem{}
Li Y.-S., De Lucia G., Helmi A., 2010, MNRAS, 401, 2036
\bibitem{}
Li Y.-S., Helmi A., De Lucia G., Stoehr F., 2009, MNRAS, 397, L87
\bibitem{}
Mac Low M.-M., Ferrara A., 1999, ApJ, 513, 142
\bibitem{}
Maeder A., 1992, A\&A, 264, 105
\bibitem{}
Marcolini A., D'Ercole A., Brighenti F., Recchi S., 2006, MNRAS, 371, 643
\bibitem{}
Marconi G., Matteucci F., Tosi M., 1994, MNRAS, 270, 35
\bibitem{}
Martin C. L., 1999, ApJ, 513, 156
\bibitem{}
Martin C. L., Kobulnicky H. A., Heckman T. M., 2002, ApJ, 574, 663
\bibitem{}
Mateo M. L., 1998, ARA\&A, 36, 435
\bibitem{}
Matteucci F., 2001, The Chemical Evolution of the Galaxy. Kluwer, Dordrecht
\bibitem{}
Matteucci F., Chiosi C., 1983, A\&A, 123, 121
\bibitem{}
Matteucci F., Greggio L., 1986, A\&A, 154, 279
\bibitem{}
Matteucci F., Recchi S., 2001, ApJ, 558, 351
\bibitem{}
Matteucci F., Tosi M., 1985, MNRAS, 217, 391
\bibitem{}
Matteucci F., Panagia N., Pipino A., Mannucci F., Recchi S., Della Valle M., 
2006, MNRAS, 372, 265
\bibitem{}
Matteucci F., Spitoni E., Recchi S., Valiante R., 2009, A\&A, 501, 531
\bibitem{}
Meurer G. R., Freeman K. C., Dopita M. A., Cacciari C., 1992, AJ, 103, 60
\bibitem{}
Mo H. J., Mao S., White S. D. M., 1998, MNRAS, 295, 319
\bibitem{}
Moster B. P., Somerville R. S., Maulbetsch C., van den Bosch F. C., Macci\`o\, 
A. V., Naab T., Oser L., 2010, ApJ, 710, 903
\bibitem{}
Mouhcine M., Contini T., 2002, A\&A, 389, 106
\bibitem{}
Nagashima M., Okamoto T., 2006, ApJ, 643, 863
\bibitem{}
Nagashima M., Lacey C. G., Okamoto T., Baugh C. M., Frenk C. S., Cole S., 2005, 
MNRAS, 363, L31
\bibitem{}
North P., et al., 2012, A\&A, 541, A45
\bibitem{}
Okamoto T., Frenk C. S., Jenkins A., Theuns T., 2010, MNRAS, 406, 208
\bibitem{}
Oppenheimer B. D., Dav{\' e} R., 2006, MNRAS, 373, 1265
\bibitem{}
Peeples M. S., Shankar F., 2011, MNRAS, 417, 2962
\bibitem{}
Pietrzy{\'n}ski G., Gieren W., Szewczyk O., Walker A., Rizzi L., Bresolin F., 
Kudritzki R.-P., Nalewajko K., Storm J., Dall'Ora M., Ivanov V., 2008, AJ, 135, 
1993
Pilkington K., et al., 2012, MNRAS, 425, 969
\bibitem{}
Pilyugin L. S., 1993, A\&A, 277, 42
\bibitem{}
Pipino A., Devriendt J. E. G., Thomas D., Silk J., Kaviraj S., 2009, A\&A, 505, 
1075
\bibitem{}
Pustilnik S. A., Kniazev A. Y., Pramskij A. G., Ugryumov A. V., Masegosa J., 
2003, A\&A, 409, 917
\bibitem{}
Putman M. E., Bureau M., Mould J. R., Staveley-Smith L., Freeman K. C., 1998, 
AJ, 115, 2345
\bibitem{}
Rahimi A., Kawata D., Allende Prieto C., Brook C. B., Gibson B. K., Kiessling 
A., 2011, MNRAS, 415, 1469
\bibitem{}
Recchi S., Hensler G., 2013, A\&A, 551, A41
\bibitem{}
Recchi S., Matteucci F., D'Ercole A., 2001, MNRAS, 322, 800
\bibitem{}
Revaz Y., Jablonka P., 2012, A\&A, 538, 82
\bibitem{}
Romano D., Cescutti G., Matteucci F., 2011, MNRAS, 418, 696
\bibitem{}
Romano D., Tosi M., Matteucci F., 2006, MNRAS, 365, 759
\bibitem{}
Romano D., Karakas A. I., Tosi M., Matteucci F., 2010a, A\&A, 522, A32
\bibitem{}
Romano D., Tosi M., Cignoni M., Matteucci F., Pancino E., Bellazzini M., 2010b, 
MNRAS, 401, 2490
\bibitem{}
Saito M., 1979, PASJ, 31, 193
\bibitem{}
Sakamoto T., Chiba M., Beers T. C., 2003, A\&A, 397, 899
\bibitem{}
Salpeter E. E., 1955, ApJ, 121, 161
\bibitem{}
Salvadori S., Ferrara A., Schneider R., 2008, MNRAS, 386, 348
\bibitem{}
Sawala T., Scannapieco C., Maio U., White S., 2010, MNRAS, 402 1599
\bibitem{}
Schaller G., Schaerer D., Meynet G., Maeder A., 1992, A\&AS, 96, 269
\bibitem{}
Schmidt M., 1963, ApJ, 137, 758
\bibitem{}
Searle L., Zinn R., 1978, ApJ, 225, 357
\bibitem{}
Shapley H., 1938, Harvard College Obs. Bulletin, 908, 1
\bibitem{}
Shetrone M. D., C{\^o}t{\'e} P., Sargent W. L. W., 2001, ApJ, 548, 592
\bibitem{}
Shetrone M., Venn K. A., Tolstoy E., Primas F., Hill V., Kaufer A., 2003, AJ, 
125, 684
\bibitem{}
Silich S. A., Tenorio-Tagle G., 1998, MNRAS, 299, 249
\bibitem{}
Smith M. C., et al., 2007, MNRAS, 379, 755
\bibitem{}
Spite M., et al., 2005, A\&A, 430, 655
\bibitem{}
Springel V., et al., 2005, Nature, 435, 629
\bibitem{}
Springel V., Wang J., Vogelsberger M., Ludlow A., Jenkins A., Helmi A., Navarro 
J. F., Frenk C. S., White S. D. M., 2008a, MNRAS, 391, 1685
\bibitem{}
Springel V., White S. D. M., Frenk C. S., Navarro J. F., Jenkins A., 
Vogelsberger M., Wang J., Ludlow A., Helmi A., 2008b, Nature, 456, 73
\bibitem{}
Springel V., White S. D. M., Tormen G., Kauffmann G., 2001, MNRAS, 328, 726
\bibitem{}
Starkenburg E., et al., 2010, A\&A, 513, A34
\bibitem{}
Starkenburg E., et al., 2013a, MNRAS, 429, 725
\bibitem{}
Starkenburg E., et al., 2013b, A\&A, 549, A88
\bibitem{}
Stil J. M., Israel F. P., 2002, A\&A, 392, 473
\bibitem{}
Summers L. K., Stevens I. R., Strickland D. K., 2001, MNRAS, 327, 385
\bibitem{}
Summers L. K., Stevens I. R., Strickland D. K., Heckman T. M., 2003, MNRAS, 
342, 690
\bibitem{}
Sutherland R. S., Dopita M. A., 1993, ApJS, 88, 253
\bibitem{}
Tafelmeyer M., et al., 2010, A\&A, 524, A58
\bibitem{}
Talbot R. J. Jr., Arnett W. D., 1973, ApJ, 186, 51
\bibitem{}
Tenorio-Tagle G., W{\"u}nsch R., Silich S., Palou{\v s} J., 2007, ApJ, 658, 1196
\bibitem{}
Thomas D., 1999, MNRAS, 306, 655
\bibitem{}
Timmes F. X., Woosley S. E., Weaver T. A., 1995, ApJS, 98, 617
\bibitem{}
Tinsley B. M., 1980, Fundam. Cosm. Phys., 5, 287
\bibitem{}
Tolstoy E., Venn K. A., Shetrone M., Primas F., Hill V., Kaufer A., Szeifert 
T., 2003, AJ, 125, 707
\bibitem{}
Tolstoy E., et al., 2004, ApJ, 617, L119
\bibitem{}
Tolstoy E., et al., 2006, Messenger, 123, 33
\bibitem{}
Tolstoy E., Hill V., Tosi M., 2009, ARA\&A, 47, 371
\bibitem{}
Vader J. P., 1987, ApJ, 317, 128
\bibitem{}
van den Hoek L. B., Groenewegen M. A. T., 1997, A\&AS, 123, 305
\bibitem{}
Veilleux S., Cecil G., Bland-Hawthorn J., 2005, ARA\&A, 43, 769
\bibitem{}
Venn K. A., et al., 2012, ApJ, 751, 102
\bibitem{}
Wang J., De Lucia G., Kitzbichler M. G., White S. D. M., 2008, MNRAS, 384, 1301
\bibitem{}
Westfall K. B., Majewski S. R., Ostheimer J. C., Frinchaboy P. M., Kunkel W. 
E., Patterson R. J., Link R., 2006, AJ, 131, 375
\bibitem{}
White S. D. M., Frenk C. S., 1991, ApJ, 379, 52
\bibitem{}
White S. D. M., Rees M. J., 1978, MNRAS, 183, 341
\bibitem{}
Wiersma R. P. C., Schaye J., Smith B. D., 2009, MNRAS, 393, 99
\bibitem{}
Wilkinson M. I., Evans N. W., 1999, MNRAS, 310, 645
\bibitem{}
Woosley S. E., Weaver T. A., 1995, ApJS, 101, 181
\bibitem{}
W{\"u}nsch R., Silich S., Palou{\v s} J., Tenorio-Tagle G., 
Mu{\~n}oz-Tu{\~n}{\'o}n C., 2011, ApJ, 740, 75
\bibitem{}
Xue X. X., et al., 2008, ApJ, 684, 1143
\bibitem{}
Yin J., Matteucci F., Vladilo G., 2011, A\&A, 531, 136
\end{thebibliography}
\end{document}